\documentclass[11pt]{article}

\usepackage{natbib}
\usepackage{graphicx}%
\usepackage{multirow}%
\usepackage{amsmath,amssymb,amsfonts}%
\usepackage{amsthm}%
\usepackage{mathrsfs}%
\usepackage[title]{appendix}%
\usepackage{xcolor}%
\usepackage{textcomp}%
\usepackage{manyfoot}%
\usepackage{booktabs}%
\usepackage{algorithm}%
\usepackage{algorithmicx}%
\usepackage{algpseudocode}%
\usepackage{listings}%

\usepackage{a4wide}
\usepackage{bbm}
\usepackage{bm}
\usepackage{subcaption}
\usepackage{algorithm}
\usepackage{algpseudocode}

\theoremstyle{definition}%
\newtheorem{example}{Example}%
\newtheorem{remark}{Remark}%

\theoremstyle{definition}%
\newtheorem{definition}{Definition}%

\newcommand{\E}{\mathbb E}
\newcommand{\e}{\mathrm e}

\newcommand{\F}{\mathcal F}

\newcommand{\bB}{\bm{\mathrm{B}}}

\newcommand{\bZ}{\bm{\mathrm{Z}}}
\newcommand{\bs}{\bm{\mathrm{s}}}
\newcommand{\Dg}{\mathrm{Dg}}
\newcommand{\D}{\mathrm{d}}

\DeclareMathOperator*{\argmax}{arg\,max}
\DeclareMathOperator*{\argmin}{arg\,min}

\begin{document}

\title{Self and mutually exciting point process embedding flexible residuals and intensity with discretely Markovian dynamics}

\author{Kyungsub Lee\footnote{Department of Statistics, Yeungnam University, Gyeongsan, Gyeongbuk 38541, Republic of Korea, Corresponding author, Email: ksublee@yu.ac.kr}}
%}

\maketitle	

%%==================================%%
%% Sample for unstructured abstract %%
%%==================================%%

\begin{abstract}
	This work introduces a self and mutually exciting point process that embeds flexible residuals and intensity with discretely Markovian dynamics. 
	By allowing the integration of diverse residual distributions, this model serves as an extension of the Hawkes process, facilitating intensity modeling.
	This model's nature enables a filtered historical simulation that more accurately incorporates the properties of the original time series. Furthermore, the process extends to multivariate models with manageable estimation and simulation implementations. We investigate the impact of a flexible residual distribution on the estimation of high-frequency financial data, comparing it with the Hawkes process.
\end{abstract}	

%%================================%%
%% Sample for structured abstract %%
%%================================%%

%%\pacs[JEL Classification]{D8, H51}

\maketitle

\section{Introduction}

This study presents a novel point process characterized by self and mutual excitation, incorporating adaptable residuals and intensity within a discretely Markovian framework.
This model can be viewed as an extension of the Hawkes process, originally introduced by \cite{Hawkes1}, which describes a stochastic intensity that is excited by events and gradually diminishes over time.
The process is known for its ability to capture the temporal clustering of events.
Initially utilized in seismology, the versatility of the process has prompted in its extensive application across diverse fields in the natural and social sciences, including stochastic financial modeling.

Early studies incorporating the Hawkes process into financial markets, such as the study by \cite{Large2007}, have explored market resilience using Hawkes processes. 
\cite{Bowsher2007} investigated the relationship between trading times and price changes in the security market.
Moreover, recent reviews by \cite{Barcry2015}, \cite{Law2015}, and \cite{Hawkes2018} 
highlighted ongoing developments in applying the Hawkes process to finance. 
This dynamic model has been extensively studied across numerous applications, emphasizing on high-frequency financial data \citep{chen2022hawkes, Morariu-Patrichi, TokeYoshida, nystrom2022hawkes, zhang2023modeling}.

This study is also motivated by the adaptation of the Hawkes process to high-frequency financial data, 
which are recorded and reported on a very precise time scale.
This data type captures price movements, trade volumes, and other market-related information with high temporal precision.
Although the residual extracted using the Hawkes model typically adheres to the standard exponential distribution, 
this characteristic may not be ensured in real-world scenarios, 
particularly in high-frequency financial markets characterized by significant noise, automated trading, and irregular behavior \citep{Lee2022}.
Although this research originated in the financial domain, we envision its extension beyond finance into various other fields.

This study proposes a process that is identical to the Hawkes model when using a unit exponential distribution but extends the framework by incorporating a more flexible residual distribution.
In contrast to the past emphasis on extending the Hawkes model in terms of the decay function \citep{Chen, Lesage}, quadratic type feedback \citep{Aubrun}, nonlinear model \citep{Gao}, multikernel \cite{Lee2023multi} and multivariate versions \citep{Lu}, 
the proposed model places emphasizes on the residual process.
By modeling the residuals, we can visualize the inferred residuals using histograms during the estimation process, allowing for a visual assessment of the model's goodness of fit. 
This approach enables model refinement and updates based on these insights. 
Although the model's definition is motivated by a flexible residual process, it ultimately leads to modeling the intensity function between events.

The proposed model shares similarities with the renewal Hawkes process presented by \cite{WHEATLEY2016120} that both models introduce a new random process to determine the arrival times. 
However, the renewal Hawkes process adapts the traditional Hawkes model by substituting a new renewal process in place of the baseline intensity. 
In contrast, the proposed model retains a baseline intensity but introduces flexibility by modifying the decaying excitation term by introducing flexible residuals.
The introduction of a flexible residual distribution means that the decay of intensity between events is not necessarily exponential. 
While non-exponential decay typically results in non-Markovian processes, which complicates simulation and estimation, our approach retains the Markov property, thereby facilitating these processes.

The proposed Hawkes model differs from the discrete-time Hawkes model suggested by \cite{browning2021simple}. 
The discrete-time Hawkes model was designed to analyze data where exact arrival times, such as the onset of coronavirus disease 2019 cases, are not available. 
This model applies a Poisson distribution to represent the number of occurrences in a pre-specified discrete time interval, allowing for multiple events in a fixed discrete period.
Conversely, our model assumes that only a single event occurs in a random discrete time interval, with the exact timing of the occurrence known. 

The enhanced model can describe a diverse range of residuals
and be effectively employed in applications such as filtered historical simulations (FHSs).  
Following the parameter estimation from the actual data, 
the residual distribution can be inferred using the model. 
The extracted residuals can be employed in an FHS, 
which is synonymous with the bootstrap method in the financial domain.
The paths generated by the FHS are expected to have probabilistic properties similar to the original time series,
compared to the Hawkes model, which uses only the unit exponential as residuals. 
This feature underscores the significance of the proposed model in comparison with its original counterpart.

Furthermore, the proposed model preserves the exponential decay properties inherent in the Hawkes model, enabling the intensity process to maintain Markov properties from a discrete-time perspective. This feature substantially simplifies the simulation and estimation procedures. 
This computational advantage is particularly notable compared to other Hawkes models that do not employ an exponential kernel and, as a result, do not exhibit Markov properties.

The remainder of this paper is organized as follows.
Section~\ref{Sect:model} introduces the process, its estimation, and simulation procedure.
Section~\ref{Sect:empirical} presents examples of estimating high-frequency financial data and measuring volatility.
Finally, section~\ref{Sect:concl} concludes this paper.

\section{Model}~\label{Sect:model}

\subsection{One-dimensional model}

Our model is one of point processes.
The point process is represented by a random countable set of points 
$\{t_1, t_2, \cdots\}$ on a measurable space $\mathcal X$ with a probability space $(\Omega, \mathcal F, \mathbb P)$.
The point process is also interpreted using a counting measure $N(A)$ 
which counts the number of points in $A \subseteq \mathcal X$.
When $\mathcal X$ denotes the timeline represented by $\mathbb R$,
$N$ can also be considered as a counting process such that
$$ N(t) = N((0,t]) = \text{the number of } t_i \text{ in }(0,t].$$
For more precise definitions, see \cite{daley2007introduction}. 

While considering the point process over $\mathbb{R}$ is mathematically convenient, in practice, processes typically start at a specific time, often $t = 0$. 
Thus, realizing point processes on the positive reals aligns better with observed data, which starts at a finite time and progresses forward.

Since our model is motivated by the Hawkes process, we recall its definition below.

\begin{definition}[Hawkes process]
	The Hawkes process $N$ is a point process on $\mathbb R$, and its
	stochastic intensity process is given by
	$$ \lambda(t) = \mu + \int_{-\infty}^{t}  h(t-s)  \D N(s)$$
	where $\mu > 0$.
	Further, $h : (0, \infty) \rightarrow [0, \infty]$ is an excitation function or a kernel. 
	The exponential function is employed in this paper such that
	$$ h(t) = \alpha \e^{-\beta t}$$
	where $0 < \alpha < \beta $. 
\end{definition}

The Hawkes process captures the self-exciting nature of events, in which past occurrences influence the likelihood of future events. 
The autoregressive form with exponential decay represents the excitation by the previous events and the diminishing intensities over time.
The intensity process is left continuous with right limits.
Given the observed event times, $\{ t_n \}$, over a finite interval $[0, T]$, the log-likelihood function of the Hawkes process  is given by the following:
$$ \sum_{t_n \leq T} \log \lambda(t_n) - \int_{0}^{T} \lambda(s) \D s.$$
The following residual of the Hawkes process follows the standard exponential distribution\footnote{In some contexts, the difference between the observed inter-arrival times and Eq.~\eqref{Eq:resid} is called the residual.}:
\begin{equation}
	\int_{t_n}^{t_{n+1}} \lambda(s) \D s. \label{Eq:resid}
\end{equation}
The inferred residual process is often employed to evaluate the goodness of fit of a model via quantile-quantile (Q-Q) plots against an exponential distribution. 
However, Q-Q plots do not perfectly align as a straight line in many applications, and this discrepancy is a central motivation for this study.

We introduce a self-exciting point process that permits a more flexible definition of the residual distribution.

\begin{definition}~\label{Def:1dH}
	A sequence of random variables, $\{ \varepsilon_n \}_{n=1}^{\infty}$, is assumed to be independent and identically distributed on a positive support with $\E[\varepsilon] = 1$,
	and $\bm{\theta}^h = \{ \mu, \alpha, \beta  \}$ denotes a parameter set with $\mu, \alpha, \beta > 0$, and $\alpha < \beta$.
	Then, stochastic processes $\{\tau_n, \lambda_n\}$ are defined such that
	\begin{align}
		&\tau_n = \phi^{-1}(\varepsilon_n; \lambda_{n-1}, \bm{\theta}^h), \label{Eq:tau}\\
		&\lambda_{n} = \psi(\tau_n; \lambda_{n-1}, \bm{\theta}^h) \label{Eq:d_lambda}
	\end{align}
	where $ \lambda_0 > \mu$ 
	and
	\begin{align}
		\psi(t; \lambda, \bm{\theta}^h) &= \mu + (\lambda - \mu + \alpha) \e^{-\beta t} \label{Eq:psi}\\
		\phi(t; \lambda, \bm{\theta}^h) &= \int_0^t \psi(s; \lambda, \bm\theta^h) \D s =
		\mu t + (\lambda - \mu + \alpha) \frac{1 - \e^{- \beta t}}{\beta}. \label{Eq:phi}
	\end{align}
\end{definition}

\begin{remark}
	Definition~\ref{Def:1dH} can be extended as follows:
	$$
	\psi(t, \lambda, \bm\theta^h) = \mu + (\lambda - \mu + \alpha) \zeta(t)
	$$
	where $ \zeta(t)$ is positive and monotonically decreases as $\zeta(t) \rightarrow 0$ as $t \rightarrow \infty $ and
	$$ \int_0^\infty \alpha \zeta (s) \D s < 1.$$
	The condition $\alpha < \beta$ is traditionally used in the Hawkes model to prevent an explosion of events in finite time. Our model also adopts this condition for stability, but further research is needed to fully explore the detailed stability criteria for our extended framework.
	In addition, the assumption that the expected value of the residuals is 1 can be relaxed to enable more versatile modeling.
\end{remark}

The following notations are used for simplicity:
$$ \psi_n(t) = \psi(t; \lambda_{n-1}, \bm{\theta}^h), \quad \phi_n(t) = \phi(t; \lambda_{n-1}, \bm{\theta}^h)$$
and where no risk of confusion occurs, the subscript $n$ may be omitted.
According to Definition~\ref{Def:1dH}, $\tau_n$ represents the inter-arrival times between events (i.e., $\tau_n = t_{n} - t_{n-1}$), and $\lambda_n$ is defined recursively through $\lambda_{n-1}$ as in the Hawkes model. 
While $\lambda_n$  may not be exactly the intensity, further explanation will be provided later.
Since $\lambda_0 > \mu$, 
by the positiveness of the parameters, $\lambda_n > \mu $ for all $n >0$
and
$$ \frac{\partial \phi(t)}{\partial t} = \psi(t) >0.$$
Hence, $\phi(t)$ is strictly monotonically increasing, 
and the inverse of $\phi$ exists.
Additionally,
$$ \E [\phi_n(\tau_n) | \F_{n-1}] = 1,$$
where $\F_{n-1}$ denotes the $\sigma$-algebra that consists of information up to time $t_{n-1}$.

If $\varepsilon$ follows a unit exponential distribution, 
Definition~\ref{Def:1dH} is equivalent to the definition of the Hawkes process,
because by Eqs.~\eqref{Eq:tau} and \eqref{Eq:phi},
\begin{align*}
	\varepsilon_n &= \phi(\tau_n; \lambda_{n-1}, \bm{\theta}^h) = \mu \tau_n + ( \lambda_{n-1} - \mu + \alpha)  \frac{1 - \e^{- \beta \tau_n}}{\beta} \\
	&= \int_{t_{n-1}}^{t_n} \mu + (\lambda_{n-1} - \mu + \alpha) \e^{-\beta s} \D s \\
	&= \int_{t_{n-1}}^{t_n} \lambda(s) \D s.
\end{align*}

Further, $\psi(t)$ in Eq.~\eqref{Eq:psi} essentially has a formula equivalent to the conditional intensity of the Hawkes process $\lambda(t)$ until the next event occurs once an event has taken place. 
This function describes how the original $\lambda$ is excited by $\alpha$ and how the part excluded by $\mu$ decays exponentially at a rate defined by $\beta$,
which can be considered a segment of the intensity function between events.
The $\phi(t)$ function is the integral of  $\psi(t)$ up to $t$. 
According to Eq.~\eqref{Eq:resid}, $\psi(\tau)$ becomes the residual. 
This work introduces the concept of applying the inverse function of $\psi$ to the residual to generate $\tau$ in reverse.

In Eq.~\eqref{Eq:d_lambda}, the $\lambda$ process is not exactly the conditional intensity in temporal point processes defined as follows:
$$\lim_{h \rightarrow 0} \frac{\E[N(t + h) - N(t) | \F_t ]}{h}.$$
To explain this,
although $\lambda_{n} $ in Eq.~\eqref{Eq:d_lambda} is defined discretely only on arrival times in the model, we consider a continuous time interval between events.
Without loss of generality, suppose that an event occurs at time 0 and the subsequent arrival, $\tau$, has not occurred by time $t$. 
Then, the conditional intensity is represented by
\begin{align}
	\lim_{h \rightarrow 0} \frac{\mathbb E[ N(t + h) - N(t)|  \mathcal F_t]}{h}  &= \lim_{h \rightarrow 0} \frac{\mathbb P[ t < \tau < t + h|   \mathcal F_t]}{h} \frac{1}{\mathbb P(\tau > t)}\\
	& =  \lim_{h \rightarrow 0} \frac{F_{\varepsilon} (\phi(t + h)) - F_{\varepsilon,t} (\phi(t ))}{h} \frac{1}{\mathbb P(\tau > t)} \\
	& = \frac{\D F_{\varepsilon}(\phi(t))}{\D t} \frac{1}{\mathbb P(\tau > t)} = \frac{ f_{\varepsilon}(\phi(t)) }{1 - F_{\varepsilon}(\phi(t))} \psi(t) \label{Eq:intensity}
\end{align}
where $f_{\varepsilon}$ and $F_{\varepsilon}$ represent the probability density function (PDF) and distribution function of $\varepsilon$, respectively.
If the residual follows the unit exponential distribution, i.e., $f_\varepsilon (t) = \e^{-t}$, then 	
$$ \frac{ f_{\varepsilon}(\phi(t))}{1 - F_{\varepsilon}(\phi(t))}  \psi(t) = \psi(t) = \mu + (\lambda(0) - \mu + \alpha) \e^{-\beta t}.$$
This is identical to the intensity function of the exponential Hawkes process from time 0 until the subsequent event.
If the residual does not follow the unit exponential distribution, the conditional intensity is not $\psi(t)$ but should be adjusted by 
$$ \frac{f_{\varepsilon}(\phi(t))}{1 - F_{\varepsilon}(\phi(t))}, $$
which depends on the distributional properties of the residual.

The above discussion highlights that while the definition of our model begins with the flexibility of applying various residual distributions, this ultimately has a direct impact on intensity modeling. 
In other words, the choice of an appropriate residual distribution can lead to the application of a suitable intensity function. 
This underscores the relationship between the residual distribution and the intensity function within our model. 
By carefully selecting the residual distribution, practitioners can effectively tailor the intensity function to better capture the dynamics of the process being modeled.

\begin{example}
	
In this example, we examine the behavior of the intensity function for the proposed point process characterized by Gamma-distributed residuals, where the intensity between events is represented by Eq.~\eqref{Eq:intensity}. 
Specifically, we consider the parameters $\mu = 0.2$, $\alpha = 0.5$, $\beta = 0.8$, $\lambda(0) = 1$, and the Gamma distribution with a shape parameter of 2 and a scale parameter of 0.5. The intensity function varies depending on the assumed distribution of the residuals. In this case, after an event occurs, the intensity starts from zero, then sharply increases, and eventually becomes nearly constant.
\end{example}

\begin{figure}[t]
	\centering
	\includegraphics[width=0.7\textwidth]{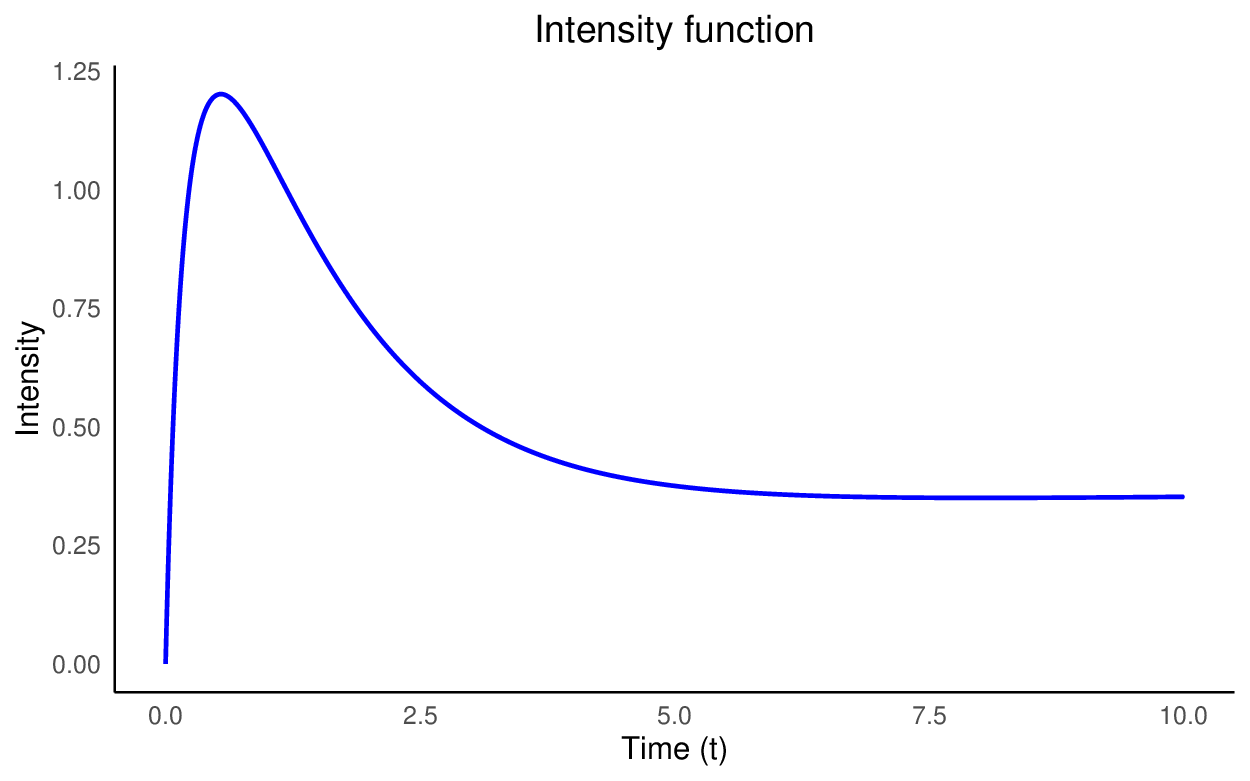}
	\caption{Intensity function between events with Gamma residual $\varepsilon \sim \Gamma(2, 0.5)$}
	\label{fig:intensity}
\end{figure}

Next, the log-likelihood function for maximum likelihood estimation (MLE) of the model is derived by
letting $f_{\varepsilon}$ be a PDF of $\varepsilon$ with the parameter set $\bm{\theta}^\varepsilon$.
By Eq.~\eqref{Eq:tau},
$$ \tau_n = \phi_n^{-1}(\varepsilon_n).$$
The conditional PDF of $\tau_n$ is represented by
$$ f_{\tau_n} (t) =  f_{\varepsilon}(\phi_n(t)) \phi'_n(t) =   f_{\varepsilon}(\phi_n(t)) \psi_n(t).$$
Thus, the log-likelihood function with the given observations of inter-arrivals $\{ \tau_n \}_{n=1}^{N}$ is as follows:
$$ \log \mathcal{L}_N (\bm \theta) =  \sum_{n=1}^{N} \ell_n(\bm\theta)$$
where $\ell_n$ (the contribution to the log-likelihood of the $n$th step with parameter set $\bm\theta = \bm{\theta}^h \cup \bm{\theta}^\varepsilon$) is
\begin{equation}
	\ell_n(\bm \theta) =  \log f_{\varepsilon}(\phi_n(\tau_n)) + \log \psi_n(\tau_n) = \log f_{\varepsilon}(\phi_n(\tau_n)) +  \log \lambda_n. \label{Eq:ln1}
\end{equation}
The maximum likelihood estimator is calculated as follows:
\begin{equation}
	\hat{\bm \theta} =  \argmax_{\bm \theta} \log \mathcal{L}_N (\bm \theta) = \argmax_{\bm \theta} \sum_{n=1}^{N} \log f_{\varepsilon}(\phi_n(\tau_n)) +  \log \psi_n(\tau_n). \label{Eq:MLE1}
\end{equation}
If $\varepsilon$ follows the standard exponential distribution, then
\begin{align}
	\ell_n(\bm \theta) &= \log \lambda_n - \phi_n(\tau_n)  = \log \lambda_n - \int_0^{\tau_n} \psi(s; \lambda_{n-1}, \bm\theta) \D s \label{Eq:llf_Hawkes}
\end{align}
which is equivalent to the log-likelihood function in the Hawkes process.

This work assumes that $\varepsilon$ does not follow the standard exponential distribution and is unknown.
For example, when fitted to the Hawkes model using inter-arrivals between events observed in high-frequency financial data, the residual often differs from the exponential distribution (see Subsection~\ref{Subsec:intraday}).
If $\bm \theta$ is estimated using Eq.~\eqref{Eq:llf_Hawkes}, which is based on the Hawkes process,
without correctly specifying the distribution of $\varepsilon$,
the estimation is a quasi-maximum likelihood estimation (QMLE).
This practice is common because the natural choice for the quasi-log-likelihood function in distributions over a semi-infinite interval is the exponential distribution.
Unlike the autoregressive conditional duration model \citep{Engle1998}, 
the QMLE of the process might be biased and inconsistent.

Consider a QMLE under the standard exponential distribution of the residual,
even if it follows a different distribution.
A sufficient condition for consistency is the martingale difference property of the score function \citep{Engle2000}.
With the true parameter $\bm{\theta}_0$, to satisfy
$$ \E[\bs_n (\bm{\theta}_0) | \F_{n-1}]  = \E\left[ \left. \frac{\partial \ell_n (\bm{\theta}_0)}{\partial \bm\theta} \right| \F_{n-1} \right] =  \bm{0},$$
that is, the score evaluated at the true parameter is a vector martingale difference sequence,
we should have
\begin{align*}
\frac{\partial}{\partial \bm{\theta}} \E[\log\lambda_n -  \phi_n(\tau_n ; \bm{\theta}_0 ) | \F_{n-1}] &=
\E \left[  \left. \frac{1}{\psi_n(\tau_n, \bm{\theta}_0)}  \frac{\partial \psi_n(\tau_n, \bm{\theta}_0)}{\partial \bm{\theta}} - \int_{0}^{\tau_n}  \frac{\partial \psi_n(s, \bm{\theta}_0) }{\partial \bm{\theta}}  \D s \right| \F_{n-1} \right]
\\
&= \bm{0}.
\end{align*}
More precisely,
\begin{align*}
	&\E \left[  \left. \frac{1}{\psi_n(\tau_n)} \e^{-\beta \tau_n} - \frac{1 - \e^{-\beta \tau_n}}{\beta} \right| \F_{n-1} \right] = 0 \\
	&\E \left[  \left. \frac{1}{\psi_n(\tau_n)} - \tau_n \right| \F_{n-1} \right] = 0 \\
	&\E \left[ \left. \frac{(1- \beta \tau_n)\e^{-\beta \tau_n}}{\psi_n(\tau_n)} - \tau_n \e^{-\beta \tau_n} \right| \F_{n-1} \right] = 0
\end{align*}
which are conditions that are too strong for the flexible distributions of $\tau$.
Therefore, to use the proposed point process, it is recommended to specify the distribution of $\varepsilon$ precisely.

What happens if we use QMLE with the exponential distribution instead of specifying the exact distribution of $\varepsilon$?
The consequences will depend on how closely the distribution of the residual aligns with the unit exponential distribution on practical application.

\begin{example}
	After simulating 50,000 paths of the proposed processes with $\mu = 0.2$, $\alpha = 0.5$, $\beta = 0.8$ and gamma-distributed residuals, we performed MLE. 
	First, we specified the likelihood function using a gamma distribution for the model.
	Second, we applied QMLE assuming the residuals follow a standard exponential distribution. 
	To ensure a gamma distribution mean of 1, we set the shape and scale parameters of the distribution equal and varied these values to study their effects. 
	Gamma distribution is the standard exponential distribution when both parameters are 1.
	The estimates are presented in Table~\ref{Tab:est}.
	The  model yields unbiased estimates, whereas the Hawkes model shows biases due to model misspecification. 
	However, these biases decrease as the residual distribution approaches an exponential form.
\end{example}

\begin{table}\caption{Comparison of estimates for the proposed and Hawkes Models}\label{Tab:est}
\begin{tabular}{c|cccc|ccc}
	\hline
	True  & \multicolumn{4}{c|}{Estimates of our model} & \multicolumn{3}{c}{Estimates of Hawkes} \\
	shape & shape & $\mu$ & $\alpha$ & $\beta$ & $\mu$ & $\alpha$ & $\beta$   \\
	\hline
	1.2 & 1.215 & 0.1993 & 0.5021 & 0.8045 & 0.2180 & 0.3763 & 0.6389 \\
	
	1.5 & 1.498 & 0.1989 & 0.5012 & 0.7965 & 0.2384 & 0.2690 & 0.4843 \\
	
	2.0 & 2.020 & 0.1991 & 0.4981 & 0.7931 & 0.2746 & 0.1610 & 0.3305 \\
	
	2.5 & 2.497 & 0.1997 & 0.4969 & 0.7958 & 0.3006 & 0.1030 & 0.2371 \\
	
	3 & 3.0545 & 0.1992 & 0.5146 & 0.8172 &  0.3328 & 0.0698 &  0.1829 \\ 
	\hline
\end{tabular}
\end{table}

Now, we propose an algorithm for the FHS,
which was introduced for the evaluation of risk under volatility models \citep{barone1998don,boudoukh1998best}, based on the bootstrap method.
FHS is utilized in risk management and volatility estimation, particularly in models such as GARCH \citep{bollerslev1986generalized}.
The use of FHS in this field aligns closely with our motivation, 
as FHS has been employed to more accurately reflect the observed residuals.
FHS provides a more realistic assessment of tail risks by using already observed extreme historical residuals to generate future paths \citep{barone2002backtesting,abad2014comprehensive,lee2018filtered,barone2018estimating}.
The algorithm is shown in Algorithm~\ref{Algo:1}.

\begin{algorithm}
	\caption{Filtered historical simulation for the one-dimensional model}\label{Algo:1}
	\begin{algorithmic}[1]
		\State \textbf{Estimate parameters:} Using the MLE with Eq.~\eqref{Eq:MLE1}, estimate $\bm{\theta}^h$ based on the observed inter-arrivals $\{ \tau_n \}_{n=1}^{N}$.
		\State \textbf{Compute }$\bm{\lambda}$\textbf{:} With estimates $\hat{\bm{\theta}}^h$, compute $\hat \lambda_n$ based on the observed $\tau_n$ and Eq.~\eqref{Eq:d_lambda}, for $n = 1, \cdots, N$, starting with some $\hat \lambda_0$.
		\State \textbf{Infer residual process:} Recall the definition of $\tau$ in Eq.~\eqref{Eq:tau}, and infer the residual process using
		\[
		\hat \varepsilon_n = \phi(\tau_n; \hat \lambda_{n-1}, \hat{\bm{\theta}}^h)
		\]
		for $n = 1, \cdots, N$.
		
		\State \textbf{Simulate inter-arrival times:} Simulate the inter-arrival times and $ \lambda$ process with some $\tilde \lambda_0$ using
		\[
		\begin{aligned}
			&\tilde \tau_{n} = \phi^{-1}(\tilde \varepsilon_{n}; \tilde \lambda_{n-1}, \hat{\bm{\theta}}^h) \\
			&\tilde \lambda_{n} = \psi(\tilde \tau_{n}; \tilde \lambda_{n-1}, \hat{\bm{\theta}}^h)
		\end{aligned}
		\]
		where $\tilde \varepsilon_{n}$ is randomly selected from $\{ \hat \varepsilon_n \}_{n=1}^{N}$ with replacement, for $n = 1, \cdots, N$.
	\end{algorithmic}
\end{algorithm}

The choice of $\hat{\lambda}_0$ in Algorithm~\ref{Algo:1} doesn't have a definitive method at present. 
However, similar to traditional Hawkes models, if the stationary conditions are satisfied, the initial value tends not to significantly impact the long-term behavior of the process after sufficient time has passed. 
Using $\hat{\mu}$ as the initial value is one of the convenient approaches.

The availability of the FHS is an advantage of the proposed model.
Based on the discussion, Algorithm~\ref{Algo:1} establishes the FHS of the process.
In the algorithm, the actual observed residuals are applied by resampling instead of assuming a specific residual distribution.

\subsection{Multivariate model}

This model can be extended to multivariate versions.
Recalling the multivariate Hawkes process, which serves as a motivational foundation for our model, we describe it through the counting processes \( N_i \) for each type  $i = 1, \cdots, m$:
$$ N_i(t) = N_i((0,t]) = \# \text{ of } t_{i,n} \in (0,t], $$
where ${ t_{i,n} }$ denotes a strictly increasing sequence of random variables with index $n$ for type $i$.
For convenience and consistency with the notation used later, $n$ is an index representing the total number of events. 
For example, $t_{1,1}$ would be the first event of type 1, $t_{2,2}$ would be the second event of type 2, and $t_{1,3}$ would be the third event of type 1 again, indicating their positions in the overall sequence.

\begin{definition}[Multivariate Hawkes process]~\label{Def:mHawkes}
	With vector representations, the multivariate Hawkes process and $\lambda$ processes are respectively defined as follows:
	$$ \bm{N}_t = \begin{bmatrix} N_1(t) \\ \vdots \\ N_m(t)  \end{bmatrix}, \quad \bm{\lambda}_t = \begin{bmatrix} \lambda_1(t) \\ \vdots \\ \lambda_m(t) \end{bmatrix}, $$
	where
	$$ \bm{\lambda}_t = \bm{\mu} + \int_{-\infty}^{t} \bm{h}(t-u) \D \bm{N}_u $$
	with constant $m \times 1$ vector $\bm{\mu} = \begin{bmatrix} \mu_1 \cdots \mu_m \end{bmatrix}^{\top}$, and
	$\bm{h}$ is defined as follows:
	$$ 
	\bm{h}(t) = \bm{\alpha} \circ 
	\begin{bmatrix} 
		\e^{- \beta_1 t} & \cdots & \e^{-\beta_1 t} \\
		\vdots & \ddots & \\
		\e^{- \beta_m t} & \cdots & \e^{-\beta_m t}
	\end{bmatrix}
	$$
	where $\bm{\alpha}$ represents a constant $m \times m$ matrix, and $\circ$ denotes the Hadamard product.
\end{definition}
For convenience, the parameter sets are defined as follows:
$$ \bm{\theta}_{ij}^h = \{ \mu_i, \alpha_{ij}, \beta_i \}, \quad \bm{\theta}_{i}^h = \{ \mu_i, \alpha_{ij}, \beta_i : 1 \leq j \leq m\}, \quad \bm{\theta}^h = \bigcup_{i=1}^{m} \bm{\theta}_{i}^h$$
Our model's multivariate version is defined in a similar manner.

\begin{definition}~\label{Def:mdH}
	The sequences of random variables, $\{ \varepsilon_{i,n} \}_{n=1}^{\infty}$, are assumed to be independent and identically distributed on a positive support with $\E[\varepsilon_i] = 1$ for $i=1,\cdots, m$.
	The stochastic processes $\{\tau_{n}, z_{n}, \lambda_{i,n}\}$ are defined such that
	\begin{align}
		\tau_{i, n} &= \phi^{-1}(\varepsilon_{i,n}; \lambda_{i, n-1}, \bm{\theta}_{ij}^h)\\
		\tau_n &= \min_{i}  \tau_{i, n}, \label{Eq:taun}\\
		z_n &= \argmin_{i} \tau_{i, n},  \\
		\varepsilon_n &= \varepsilon_{z_n, n},  \\
		\lambda_{i,n}& = \psi(\tau_{n}; \lambda_{i,n-1}, \bm{\theta}_{ij}^h)  \label{Eq:mlambda}
	\end{align}
	with $j = z_{n-1}$ (i.e., the type of event in the previous step) and  $ \lambda_{i,0} > \mu_i$. 
	In addition,
	\begin{align}
		\psi(t; \lambda, \bm{\theta}_{ij}^h) &= \mu_i + (\lambda - \mu_i + \alpha_{ij}) \e^{-\beta_i t} \\
		\phi(t; \lambda, \bm{\theta}_{ij}^h) &= \int_0^t \psi(s; \lambda, \bm{\theta}_{ij}^h) \D s =
		\mu_i t + (\lambda - \mu_i + \alpha_{ij}) \frac{1 - \e^{- \beta_i t}}{\beta_i}.
	\end{align}
\end{definition}

The central concept involves identifying the minimum value among the $\tau_{i, n}$ terms in Eq.~\eqref{Eq:taun}, aligning with the Hawkes simulation technique employed by \cite{dassios2013exact}.
This approach is another widely used method in Hawkes simulations, which are distinct from other popular thinning algorithms \citep{ogata1981lewis}.
If $\varepsilon$ follows a unit exponential distribution, the multivariate model is also equivalent to the typical multivariate Hawkes model.

For simplicity,
$$ \psi_{i,n}(t) = \psi(t; \lambda_{i, n-1}, \bm{\theta}_{i z_{n-1}}^h), \quad \phi_{i,n}(t) = \phi(t; \lambda_{i, n-1}, \bm{\theta}_{i z_{n-1}}^h).$$
For the multivariate version, the $n$-th contribution of the log-likelihood function $\ell_n$ is
\begin{equation}
	\ell_n(\bm{\theta}) = \log \psi_{z_n, n}(\tau_{n}) + \log f_\varepsilon (\phi_{z_n, n}(\tau_n)) + \sum_{i\neq z_n}\log ( 1 - F_{\varepsilon}(\phi_{i, n}(\tau_n))) \label{Eq:llf2}
\end{equation}
where $F_{\varepsilon}$ denotes the cumulative distribution function of $\varepsilon$
and the maximum likelihood estimates are
\begin{equation}
	\hat{\bm \theta} = \argmax_{\bm \theta} \sum_{n=1}^{N} \ell_n(\bm{\theta}). \label{Eq:MLE2}
\end{equation}
Let $n_i \subseteq \{1, \cdots, N \}$ denote the ordered sequence of indices associated with type $i$
and let $n_i(k)$ denote the $k$-th element of $n_i$. 
Then, the residual process associated with $i$, $\varepsilon_{i,k}$, is defined by
\begin{equation}
	\varepsilon_{i,k} = \sum_{n = n_i(k) + 1}^{n_i(k+1)} \phi_{i} (\tau_n; \lambda_{i,n-1}, \bm{\theta}_{iz_{n-1}}^h ).\label{Eq:mres}
\end{equation}
Algorithm~\ref{Algo:2} establishes the simulation method of the multivariate process.
While the multivariate nature makes it challenging to apply the semi-parametric bootstrap as in the one-dimensional case, the simulation can still be performed by generating $\varepsilon$ using the estimated parameters.
 
\begin{algorithm}
	\caption{Simulation for multivariate model}\label{Algo:2}
	\begin{algorithmic}[1]
		\State \textbf{Estimate parameters :}  With the given observations, $\{\tau_n, z_n\}_{n=1}^{N}$, and using the MLE with Eq.~\eqref{Eq:MLE2}, estimate $\bm{\theta}$.
		
		\State \textbf{Compute }$\bm{\lambda}$\textbf{:}  With $\{\tau_n, z_n\}_{n=1}^{N}$ and estimates $\hat{\bm{\theta}}^h$ obtained in Step 1, compute $\hat \lambda_{i,n}$ using Eq.~\eqref{Eq:mlambda} with some $\hat \lambda_{i,0}$ for each $i$.
		
		\State \textbf{Generate residual processes:} Generate the residual processes $\tilde \varepsilon_{i, n}$ for each type $i$ using the estimates $\hat{\bm{\theta}}$ under the distribution $F_\varepsilon$. 
		
		\State \textbf{Simulate inter-arrival times:} Simulate the inter-arrival time, type, and  $\lambda$ processes as follows:
		\[
		\begin{aligned}
			&\tilde \tau_{i,n} = \phi^{-1}(\tilde \varepsilon_{i,n}; \tilde \lambda_{i,n-1}, \hat{\bm\theta}^h) \\
			&\tilde \tau_n = \min_{i} \tilde \tau_{i, n}, \\
			&\tilde z_n = \argmin_{i} \tilde \tau_{i, n}, \\
			&\tilde \lambda_{i,n} = \psi(\tilde \tau_{n}; \tilde \lambda_{i,n-1}, \hat{\bm\theta}^h)
		\end{aligned}
		\]
		for some $\tilde \lambda_{i,0}$.
	\end{algorithmic}
\end{algorithm}

Although the Hawkes model has numerous favorable properties, it occasionally fails to fit real-world data adequately. 
The advantage of the our model is its ability to introduce a flexible residual process while maintaining an exponential decay structure, enabling it to achieve high goodness of fit across diverse datasets. 
The following section explores examples.

\section{Empirical example}~\label{Sect:empirical}

\subsection{Intraday financial data}\label{Subsec:intraday}

\subsubsection{Data}

In this subsection, we apply the proposed model using ultra-high-frequency financial data. 
Nowadays, major stock exchanges publish intraday data on asset price dynamics and market transactions with timestamps recorded with an ultra-high-frequency time resolution.
For instance, the limit order information for stocks listed on the New York Stock Exchange from the Consolidated Tape System has evolved in terms of time resolution: from milliseconds up to July 2015, to microseconds up to September 2018, and nanoseconds afterward.
 
These data encompass details, including price quotes, price changes, trade volumes, bid-ask spreads, limit and market order arrivals, transactions, and cancellations. 
Major stocks with active trading can generate tens to hundreds of thousands of trade-related records in a single day.
The availability and analysis of high-frequency data have become increasingly critical in financial markets due to advancements in technology and trading strategies, 
allowing market participants to react quickly to changing market conditions, execute trades with precision timing, and implement algorithmic trading strategies.

The empirical study uses the mid-price of AAPL (Apple Inc.)'s national best bid and offer on a randomly chosen date: November 7, 2019. 
The mid-price (the average of the best bid and ask prices) is selected because it tends to be more stable and less noisy, mitigating the effects of the bid-ask bounce compared to transaction prices. 
This choice is consistent with established practices in the literature, for example, in work by \cite{Bacry2014}, \cite{palguna2016mid}, \cite{wehrli2021scale}, and \cite{magris2023bayesian}.
This dataset comprises approximately 300,000 records related to changes in the best bid or ask price and outstanding limit order volumes.

This work focuses on data from 10:00 to 15:30, encompassing regular trading hours with a 30-minute margin after the market opens and before it closes. 
The arrival times are defined as the moments when the mid-price changes. 
Thus, there are 66,258 observations which is fewer than the original dataset but still substantial.

\subsubsection{Residual of the Hawkes model}

First, a one-dimensional standard exponential Hawkes model was estimated with the arrival times of mid-price changes, regardless of their direction.
The maximum likelihood estimates under the one-dimensional standard Hawkes model are
\begin{equation}
\hat \mu = 1.477 (0.0033), \enspace \hat \alpha = 799.1(0.0134),\enspace \hat \beta = 1428(0.0136) \label{Eq:est_MLE}
\end{equation}
with the numerically estimated standard errors in parentheses.

\begin{figure}
	\centering
	\includegraphics[width=0.5\textwidth]{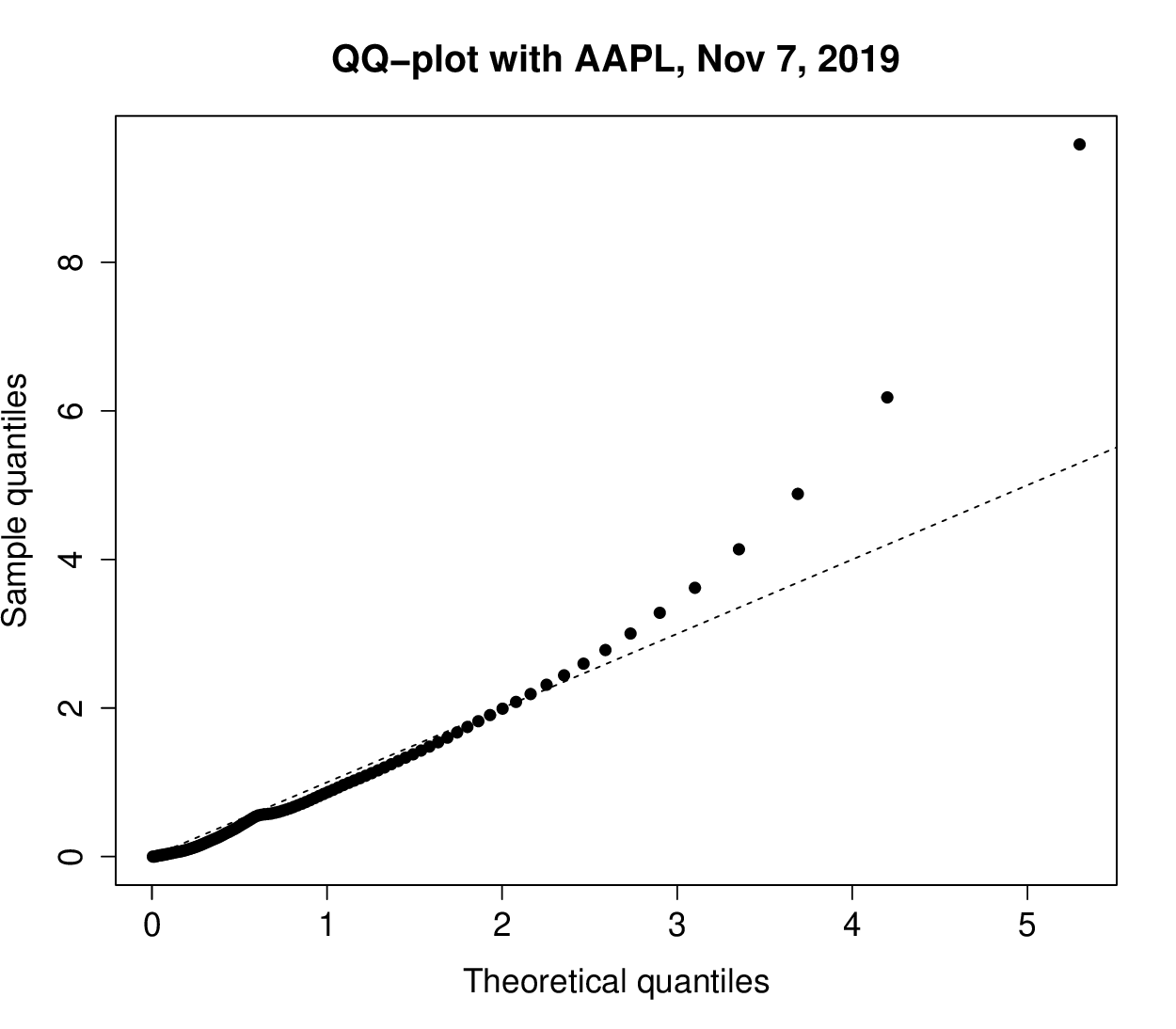} 
	\caption{Quantile-quantile plot of residuals under the standard Hawkes process estimated using high-frequency financial data }\label{Fig:QQ}
\end{figure}

The ultra-high-frequency activity driven by numerous market participants and financial firms equipped with automated transaction technology results in many events in a very short time interval, leading to significantly high estimates.
Figure~\ref{Fig:QQ} assess how well this estimate fits, using a Q-Q plot of the inferred residuals defined by Eq.~\eqref{Eq:resid} with inferred $\hat \lambda$ instead of $\lambda$.
As illustrated in the plot, the fit is inaccurate in the tail regions.
One method to address this problem under the Hawkes framework is applying a multikernel model, as in \cite{Lee2023multi}.
However, this approach increases the model complexity.

\begin{figure}
	\centering
	\begin{subfigure}[b]{0.47\textwidth}
		\centering
		\includegraphics[width=\textwidth]{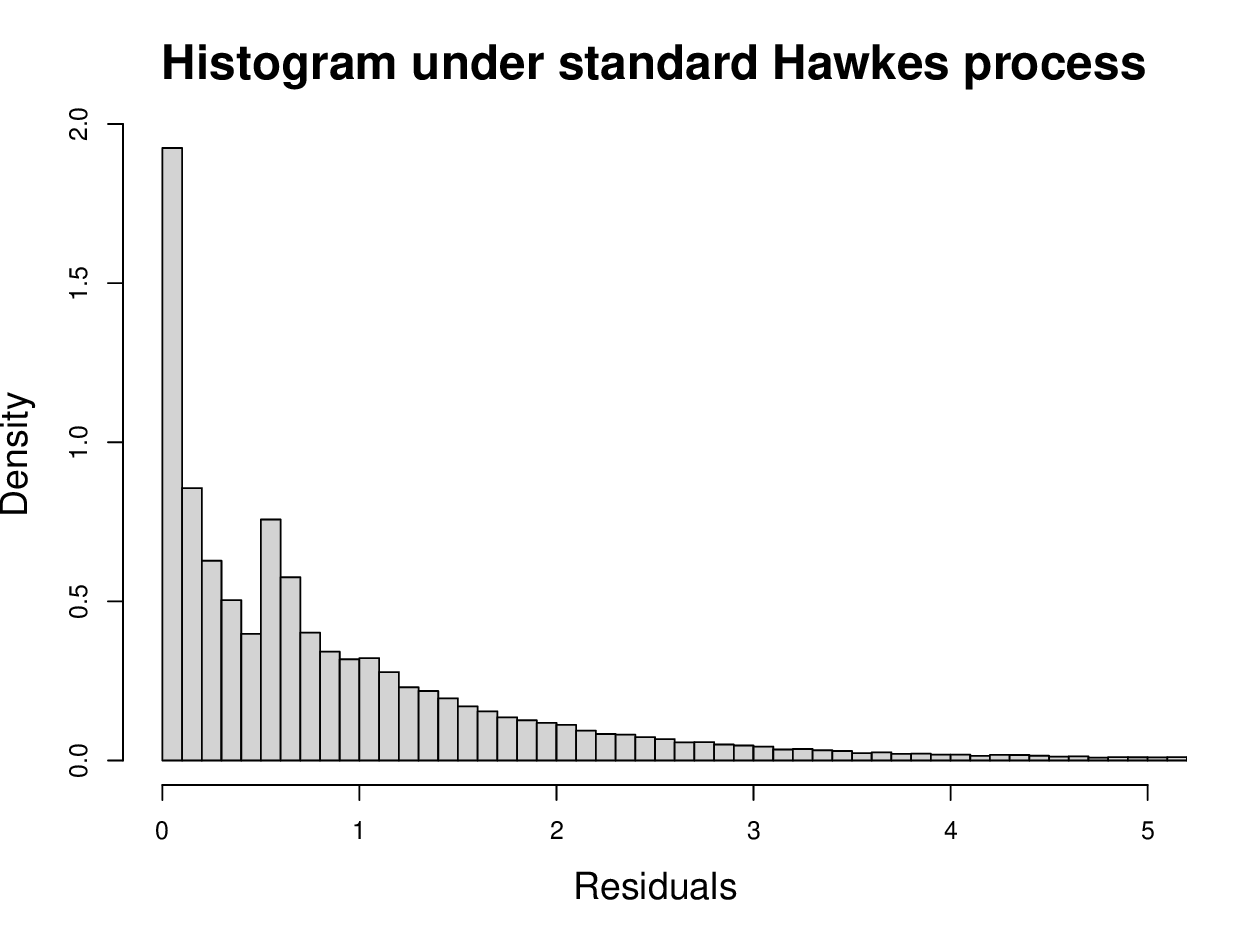}
		\caption{Hawkes process residuals}
	\end{subfigure}
	\quad
	\begin{subfigure}[b]{0.47\textwidth}
		\centering
		\includegraphics[width=\textwidth]{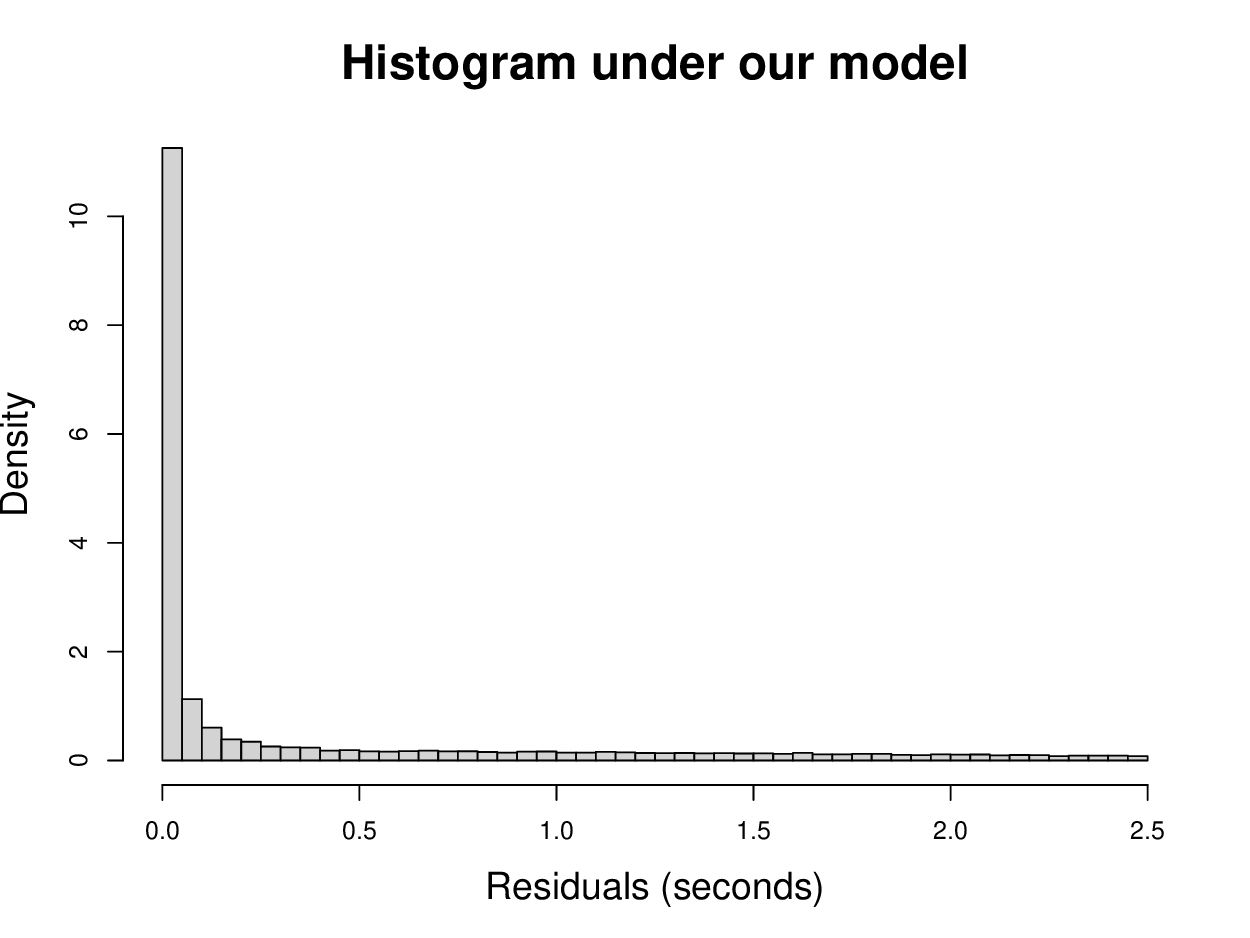}
		\caption{Our model residuals}
	\end{subfigure}
	\caption{Histograms of residuals inferred under the Hawkes process (left) and our model (right) based on AAPL data from November 7, 2019.}
	\label{Fig:residual1}
\end{figure}

The histogram on the left side of Figure~\ref{Fig:residual1} displays the inferred residuals under the standard Hawkes model.
More events occur in both tails than expected in the unit exponential distribution. 
Events that correspond to extremely small inferred residuals are more likely to occur than those in the unit exponential distribution.
A notable concentration of residuals is observed close to zero, indicating an ultra-high frequency of events occurring in very short intervals in the stock market.

Events associated with large residuals are also more frequent than in the standard exponential distribution. 
Although omitted from the figure, residuals greater than five account for 2.4\% of the total, significantly higher than the 0.7\% expected from the unit exponential distribution.

The MLE under the standard Hawkes process aims to determine the best fit of the distribution of the inferred residuals to the unit exponential distribution by adjusting the parameter values. 
The next step is observing what happens when more flexibility is gained by removing the exponential distribution constraint.

\subsubsection{Residual of our model}

This study employs the generalized method of moments (GMM)
to examine the residual distribution more flexibly.
This technique is widely used in econometrics and statistics due to its versatility and robustness \citet{hansen1982large}. 
Based on our empirical study experience, although MLE is typically recommended for the proposed model owing to its relative numerical stability, alternative methods may be required when the mathematical formula for the residual density function is not well understood. 
In such cases, a semi-parametric approach, such as the GMM or the generalized empirical likelihood method \citep{owen1990empirical, newey2004higher}, could provide a helpful starting point for determining the shape of the residual distribution.

The moment conditions are given by the following:
\[
\mathbb{E}[\mathbf{g}(\bm{\theta}^h)] = \mathbb{E}
\begin{bmatrix}
	\varepsilon_{n-1} (\varepsilon_n - 1) \\
	\lambda_{n-1} (\varepsilon_n - 1) \\
	\tau_{n-1} (\varepsilon_n - 1) \\
\end{bmatrix} = \mathbf{0}
\]
where we use the fact that $\mathbb{E}[\varepsilon_n] = 1$ and that $\varepsilon$ is independent of the lagged variables of $\varepsilon, \lambda, \tau$.
Thus, the corresponding GMM estimator
is defined as follows:
$$\argmin_{\bm{\theta}^h} \overline{\mathbf{g}}^{\top} \mathbf{W} \overline{\mathbf{g}},$$
where $\overline{\mathbf{g}}$ denotes to the sample mean of $\mathbf{g}$
computed using the inferred residual $\hat \varepsilon$ and $\hat \lambda$
with a positive semi-definite matrix $W$.
This method has been employed as an estimation approach in the field of financial econometrics, as demonstrated in the work by \cite{bollerslev2011dynamic, garcia2011estimation, lee2016probabilistic}.
Although many aspects of the GMM approach for the proposed model still require further research, such as the optimal equality conditions and numerical stability, 
these problems are beyond the scope of this paper.

A two-step GMM approach was employed to enhance the efficiency of the estimator by updating the weighting matrix $\mathbf{W}$. 
Initially, the approach start with the identity matrix and updates it to the inverse of the sample variance-covariance matrix of $\mathbf{g}$.
The estimates are
\begin{equation}
\hat{\mu} = 2.023(0.0779), \quad \hat{\alpha} = 9.606(6.347), \quad \hat{\beta} = 19.36(12.99), \label{Eq:est_GMM}
\end{equation}
where the numerically computed standard errors are provided in parentheses.
The estimates in \eqref{Eq:est_GMM} are different from \eqref{Eq:est_MLE}, but the branching ratios, $\alpha / \beta$, are similar.
 
The inferred residual process is calculated using these estimates, and the histogram of the residual is displayed on the right in Figure~\ref{Fig:residual1}. 
Compared with the residuals from the standard Hawkes process, the residual distributions from the suggested model have heavier tails. 
The concentration of residuals at close to zero is more pronounced than the histogram from the standard Hawkes process, 
and the proportion of residuals larger than five is approximately 5\% which is greater than the 2\% occurring in the standard Hawkes process.
Although this study recommends the MLE to estimate the proposed model due to its numerical stability in this context, 
the extreme distribution of residuals and the absence of an appropriate mathematical formula for the residual density function require GMM in this example.

Comparing these two results demonstrates that the interpretations of each model differ. 
The standard Hawkes model, which employs the basic exponential distribution for residuals, 
interprets high-frequency activity as having higher intensity, stronger excitation, and fast dissipation per event. 
In contrast, our model fits the data with a more extreme residual distribution,
implying that the ultra-high frequency activities are owing to a fat-tailed residual distribution.
Thus, $\alpha$, the size of the excitation, and $\beta$, representing the decay, have lower values in our model.

In the proposed model, the introduction of complexity modifies the interpretation of parameters such as $\alpha$ and $\beta$ when compared to the original Hawkes model. 
We focus more on improving how well the model fits the data and predicts outcomes, 
albeit potentially at the expense of the straightforward interpretability of these parameters.
The flexible residual distribution helps us fit the data better, but it implies that conventional interpretations related to the excitation size $\alpha$ and decay rate $\beta$ may not directly apply.

The residual distribution in our framework is closely linked to the actual distribution of inter-arrival times. 
If the inter-arrival distribution has fat-tailed characteristics, the residual distribution is likely to show similar fat-tailed properties. 
Our model then aims to minimize the influence of preceding events by using the excitation and decay structure, making the residual distribution as independent as possible. 
While knowing that the residuals have a specific shape, such as a fat tail, might not have direct applications, it is crucial for deriving a more accurate inter-arrival distribution. 
This understanding enhances our ability to predict durations, which can be critical for the model's application.

%여기부터

\subsubsection{Simulation}

\begin{figure}
	\centering
	\includegraphics[width=0.47\textwidth]{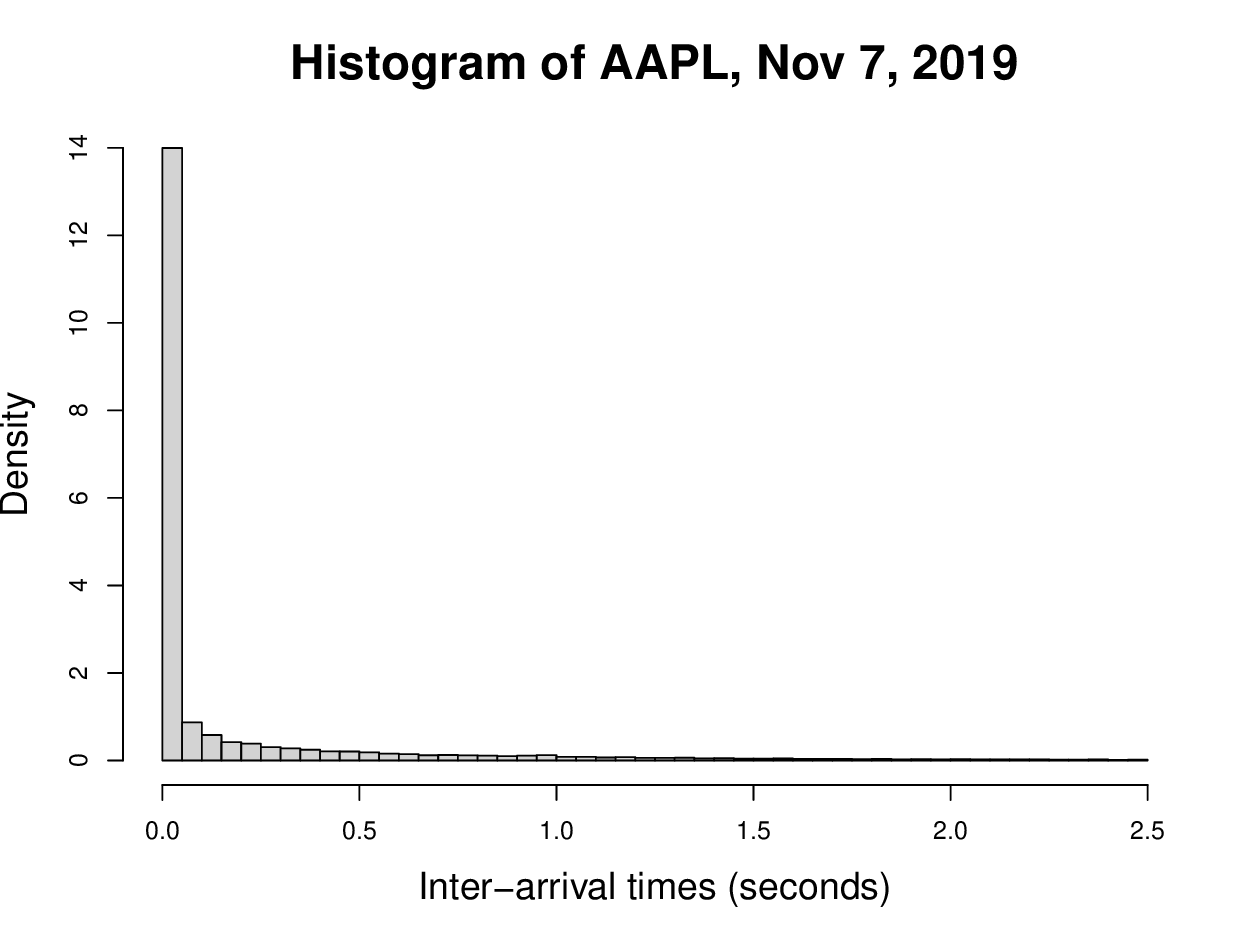}
	\caption{Histogram of inter-arrival times of AAPL on November 7, 2019}\label{hist:AAPL}
\end{figure}

\begin{figure}
	\begin{subfigure}{0.47\textwidth}
		\centering	
		\includegraphics[width=\textwidth]{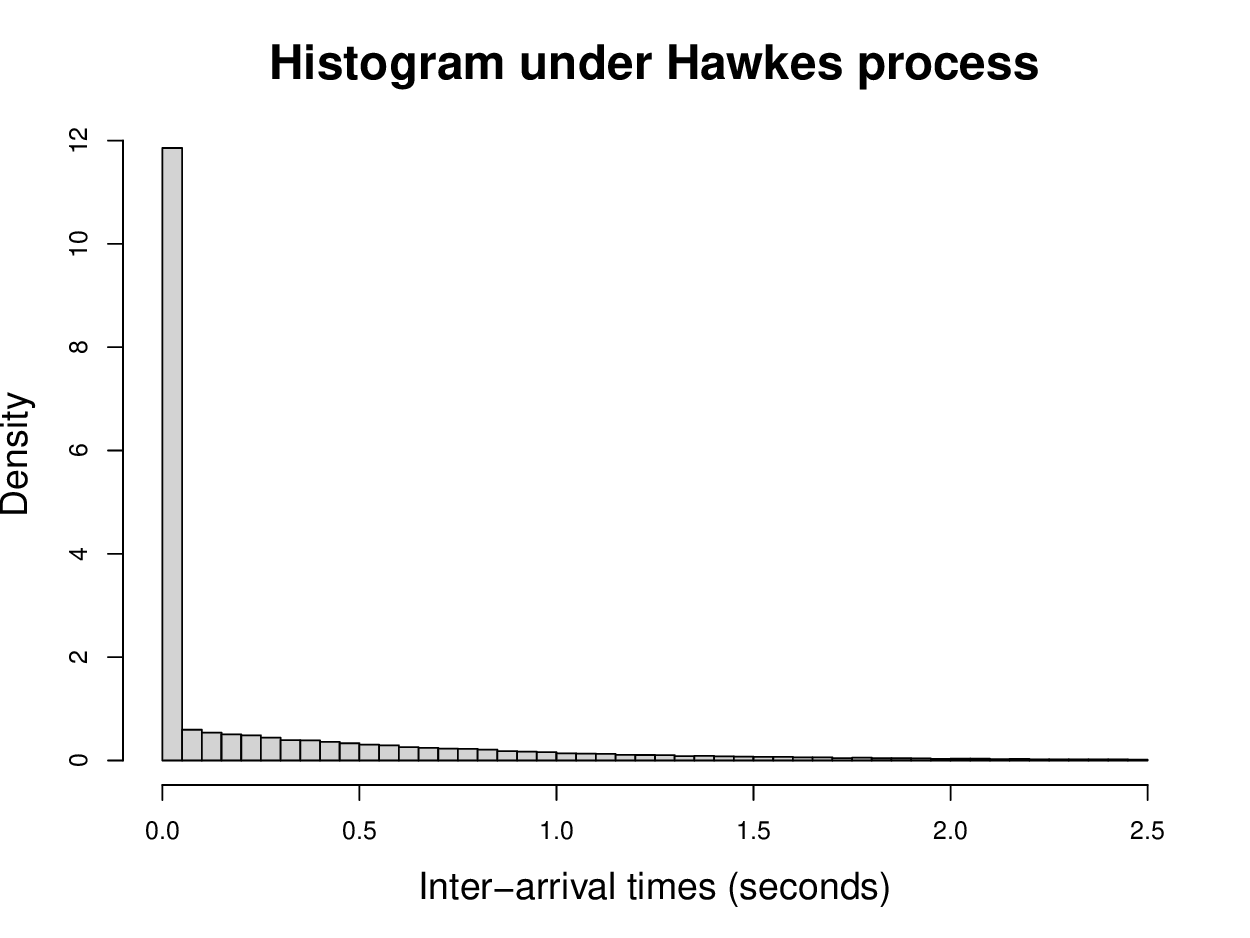}
		\caption{Hawkes process}\label{hist:Hawkes}
	\end{subfigure}
	\quad
	\begin{subfigure}{0.47\textwidth}
		\centering
		\includegraphics[width=\textwidth]{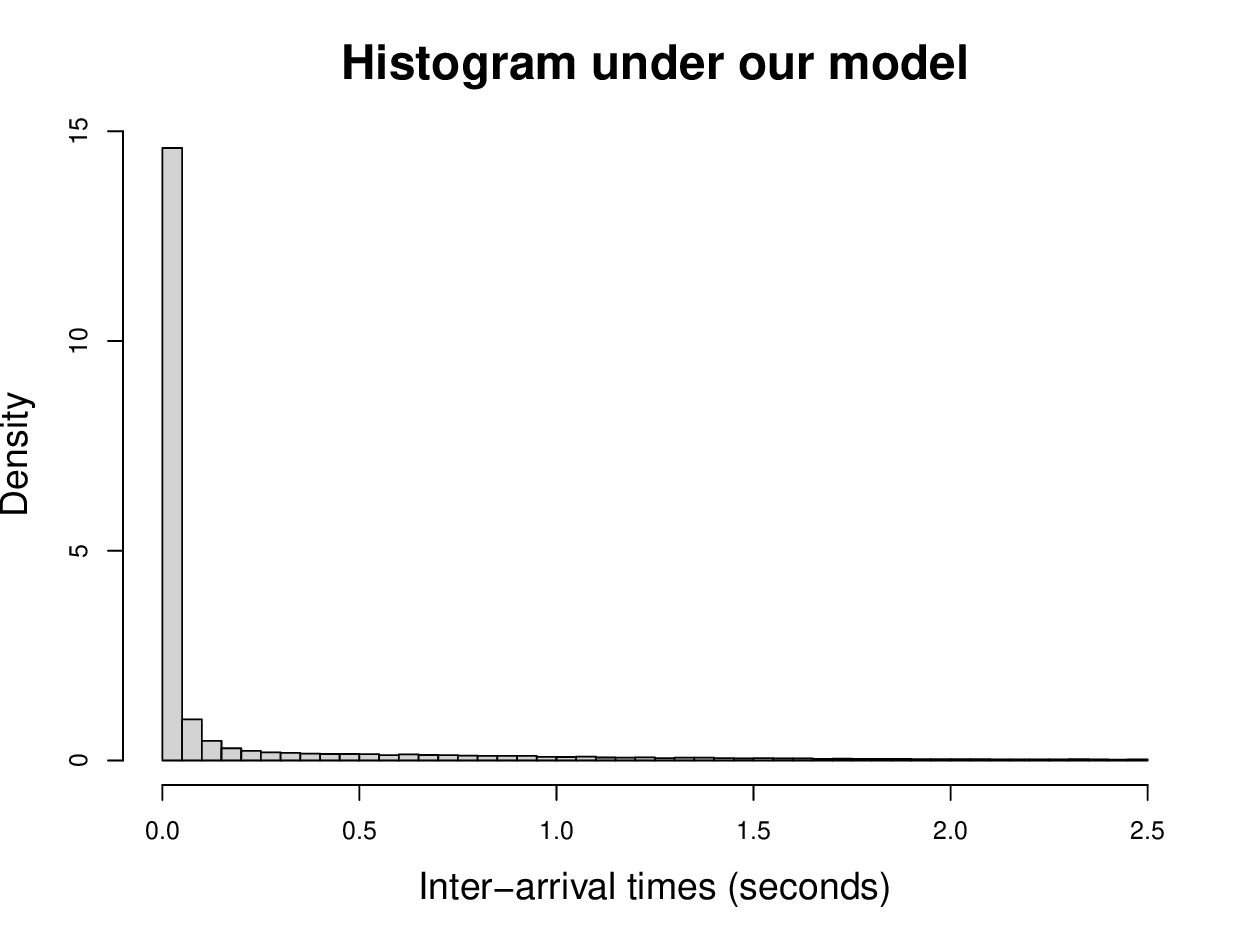}
		\caption{Our model}\label{hist:dHawkes}
	\end{subfigure}
	\caption{Histograms of inter-arrival times simulated under the Hawkes process (left) and our model (right)}
	\label{Fig:hist_arrivals}
\end{figure}

An FHS was performed with these estimates to compare the distributions of the inter-arrival times. 
First, Figure~\ref{hist:AAPL} displays the histogram of the actual inter-arrival times of the mid-price changes of AAPL from November 7, 2019.
Next, the same number of arrival events as the actual events are generated using the MLE estimates in Eq.~\eqref{Eq:est_MLE} and the standard Hawkes model.
Figure~\ref{hist:Hawkes} illustrates the histogram of the inter-arrival times from this simulation. 
Next, the FHS introduced in Algorithm~\ref{Algo:1} is conducted using the estimates in Eq.~\eqref{Eq:est_GMM} 
and the residuals from the GMM procedure under our model. 
Figure~\ref{hist:dHawkes} presents the histogram of the inter-arrival times from this simulation.

Although the three histograms may initially appear similar, 
the proposed model outperforms the standard Hawkes model  in fitting extreme distributions.
For instance, inter-arrival times of less than 0.1 milliseconds account for 20.5\% of the actual observed AAPL data, with our model estimating 19.4\% and the standard Hawkes model estimating only 8.9\%. 
Inter-arrival times greater than 3 seconds constitute 2.0\% of the actual observations, with our model estimating 2.2\% compared to just 0.5\% estimated by the standard Hawkes model.

\subsection{Sparsely observed data}

\subsubsection{Data preprocessing}

Despite the advantages of high-frequency financial data, 
challenges include data cleaning, noise filtering, and computational intensity due to the sheer volume of generated data \citep{Andersen2000}. 
Owing to the microstructure noise in raw data, 
sparsely modified data can be employed for financial quantitative analysis instead of raw data.
An example is the Hawkes volatility derived from the parameters in $\bm{\theta}^h$ based on sparsely observed high-frequency financial price data, which demonstrates promising performance \citep{LeeHawkesVol}.
There may be numerous methods to generate sparse data, but this study applies the method from \cite{LeeHawkesVol}.
This example demonstrates the ability of the suggested model to capture the irregular shape of the residuals.

Let $P(t)$ denote the mid-price at time $t$.
In addition, $\Delta t$ represents a fixed size for sparse modification,
but the price at multiples of $\Delta t$ are not automatically recorded.
Instead, the price change is recorded at more flexible time using with the following procedure 
to describe the random arrivals.

Starting with time 0 and $P(0)$, we observe $P(\Delta t)$. 
If $P(\Delta) \neq P(0)$, we determine the last time $t < \Delta t$,
when the price changed to $P(\Delta t)$.
This time $t$ and price $P(\Delta t)$ are recorded as sparsely modified data.
The procedure is repeated for each $n\Delta t$ and $P(n \Delta t)$,
i.e., finding the last time before $n\Delta t$ where the price is equal to $P(n \Delta t)$.

During the procedure, if $P(n \Delta t)$ is the same as the previously recorded price, 
i.e., there was no price change,
the procedure moves to the next step without recording it.
This method is designed to ensure the randomness of the arrival times on a continuous domain while selecting the data.
For example, data is selected over only fixed discrete time intervals, the arrival times occur only as multiples of the constant $\Delta t$, 
differing from the typical Hawkes model
where the arrival times are defined on a continuous domain.

The interval length $\Delta t$ is subjectively determined  
because the optimal interval is unknown.
This subsection empolyes the sparsely observed stock price of AAPL on  November 7, 2019, with an interval length of 0.5 seconds.
This subsection discusses the estimation and simulation results and
examines the effect of the interval length using the proposed model and FHSs.

\subsubsection{One-dimensional model}\label{subsubsec:oned}

First, the standard Hawkes model was applied to the sparsely observed mid-price changes as in Subsection~\ref{Subsec:intraday}.
The following estimates are obtained:
\begin{equation}
	\hat\mu = 0.0389(0.0159), \quad \hat\alpha= 0.0154(0.0013), \quad \hat\beta = 0.0160(0.0014). \label{Eq:est_eHawkes}
\end{equation}
Compared to the estimates in \eqref{Eq:est_MLE}, the values of the current estimates are significantly smaller. 
This result stems from omitting numerous ultra-high-frequency noises during the data preprocessing.

\begin{figure}
	\centering
	\begin{subfigure}[b]{0.45\textwidth}
		\centering
		\includegraphics[width=\textwidth]{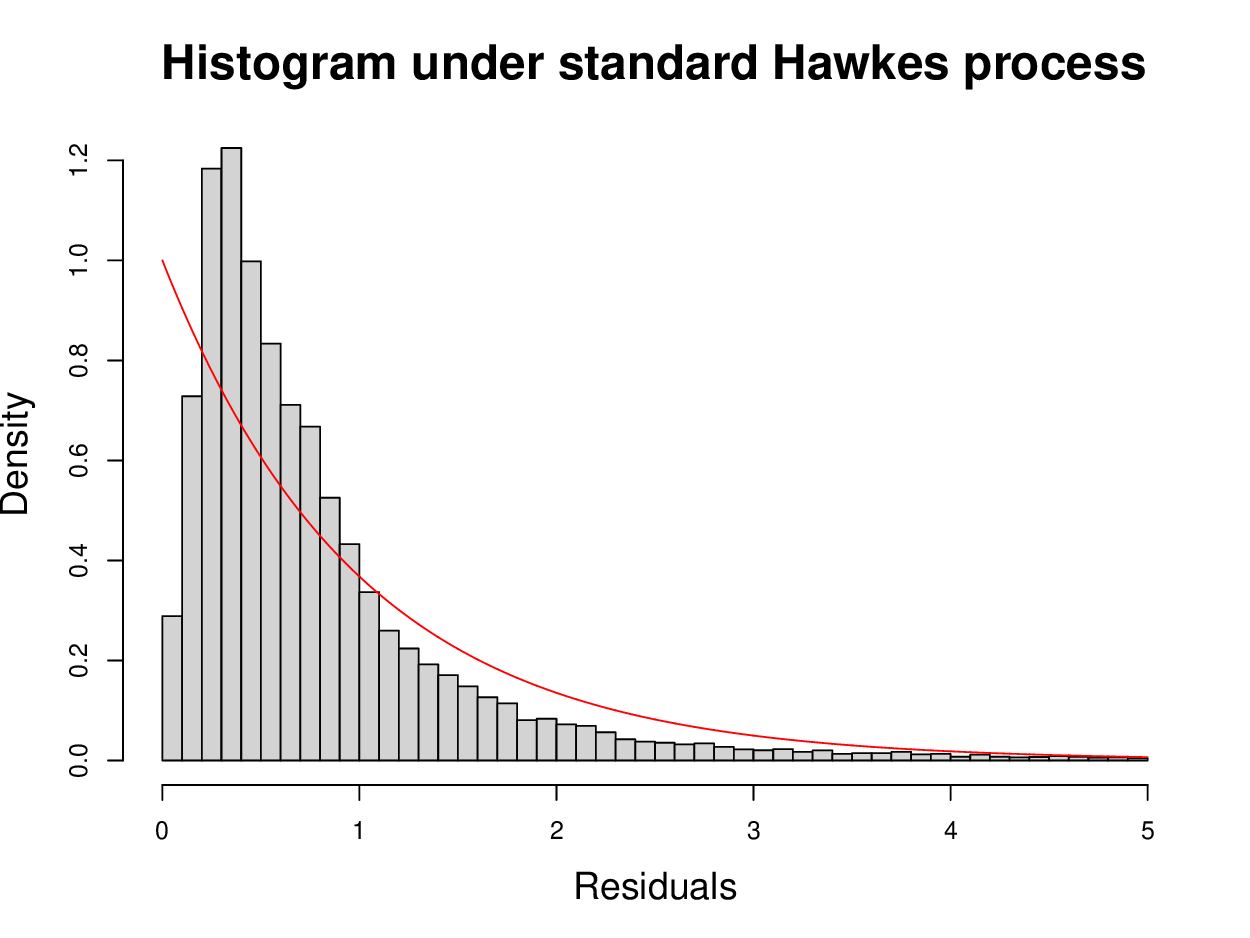}
		\caption{Hawkes process residuals}\label{Fig:residual_emhawkes}
	\end{subfigure}
	\qquad
	\begin{subfigure}[b]{0.45\textwidth}
		\centering
		\includegraphics[width=\textwidth]{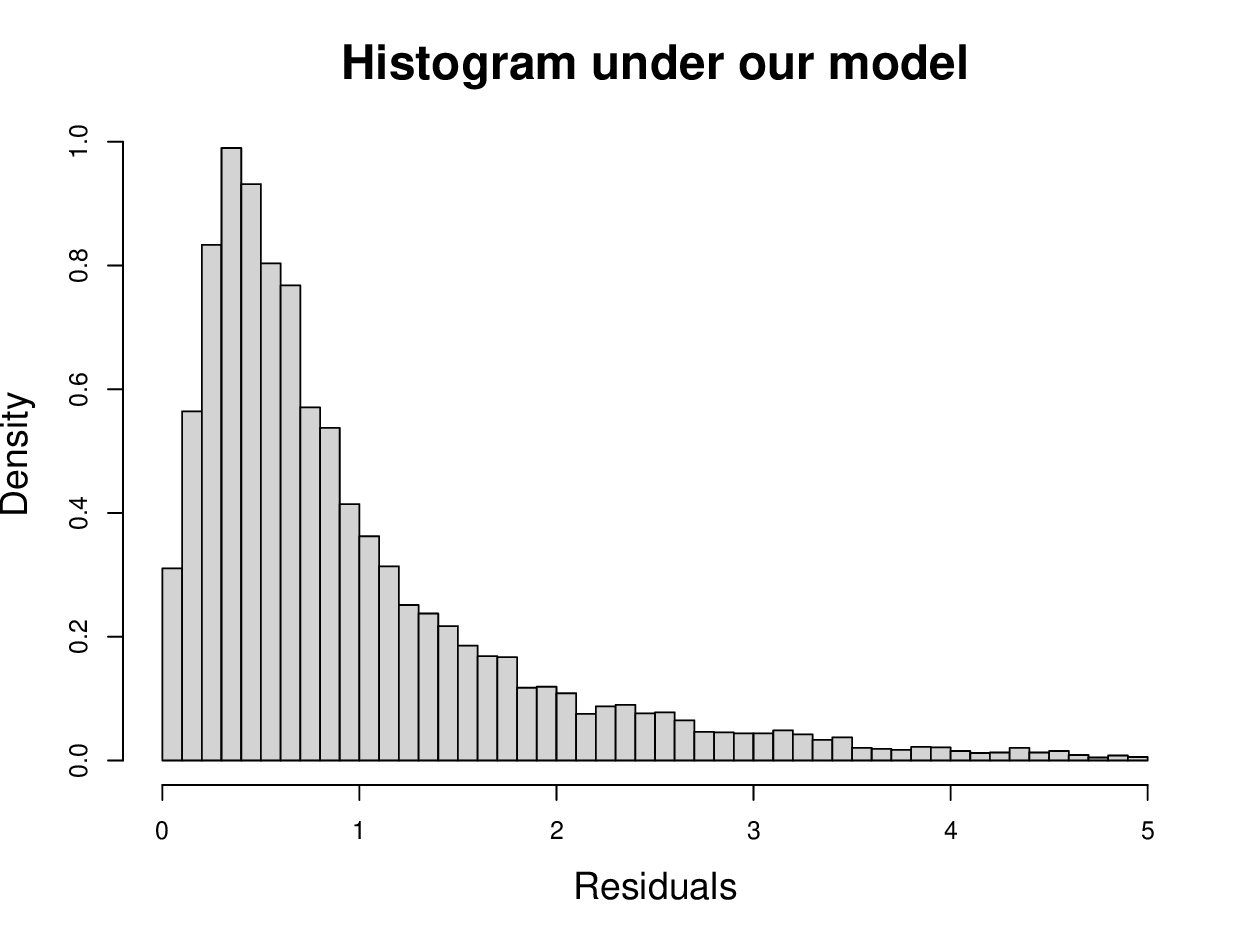}
		\caption{Our model residuals}\label{Fig:residual_dhawkes}
	\end{subfigure}
	\caption{Histograms of residuals under the Hawkes process (left) and our model (right), based on sparsely observed AAPL data with 0.5-second intervals, November 7, 2019.}
	\label{Fig:residual}
\end{figure}

Similar to the approach in Subsection~\ref{Subsec:intraday},
the residuals inferred using the estimates and the standard Hawkes model are depicted in Figure~\ref{Fig:residual_emhawkes}. 
Due to data preprocessing of the sparse observations, the extracted residuals show a peak offset from zero and do not follow the standard exponential distribution.
However, compared to the Hawkes process residuals from Subsection~\ref{Subsec:intraday}, the distribution of the residuals more closely approximates the exponential distribution.

Second, the same data were fitted to the proposed model using a more flexible PDF of $\varepsilon$ than the standard exponential, such that
\begin{equation}
f_\varepsilon(x) = 
\begin{cases} 
	\frac{p \ell - c}{a}x + c , & \text{if } 0 < x < a \\ 
	p \ell \e^{-\ell (x-a)}, & \text{if } a \leq x,
\end{cases}  \label{Eq:f}
\end{equation}
where $ c = \dfrac{2 - 2p - p a \ell}{a}$.
For $0 < x < a$, the PDF is trapezoidal with probability $1-p$, and for $a \leq x$, it is the exponential distribution of the rate $\ell$ with probability $p$.
Given that the expected value of $\varepsilon$ is 1,
$$ p = \frac{6 \ell - 2 a  \ell}{a^2\ell^2  + 4 a \ell  + 6}.$$
When $p = 1$ and $a = 0$, the distribution is standard exponential.
The density function is designed based on extensive empirical observations relevant to this case.

The resulting estimates are as follows:
\begin{align}
&\hat\mu = 0.1573(0.0178), \quad \hat\alpha = 0.0327(0.0053), \quad \hat\beta = 0.0436(0.0080), \\
&\hat a = 0.3053(0.0052), \quad \hat\ell = 1.531(0.0026)
\end{align}
Figure~\ref{Fig:residual_dhawkes} presents the histogram of residuals obtained using the flexible PDF and proposed model to fit the shape of this residual, which is depicted on the right of the figure and displays a better fitting result.

\begin{figure}[t]
	\centering
	\begin{subfigure}[b]{0.45\textwidth}
		\centering
		\includegraphics[width=\textwidth]{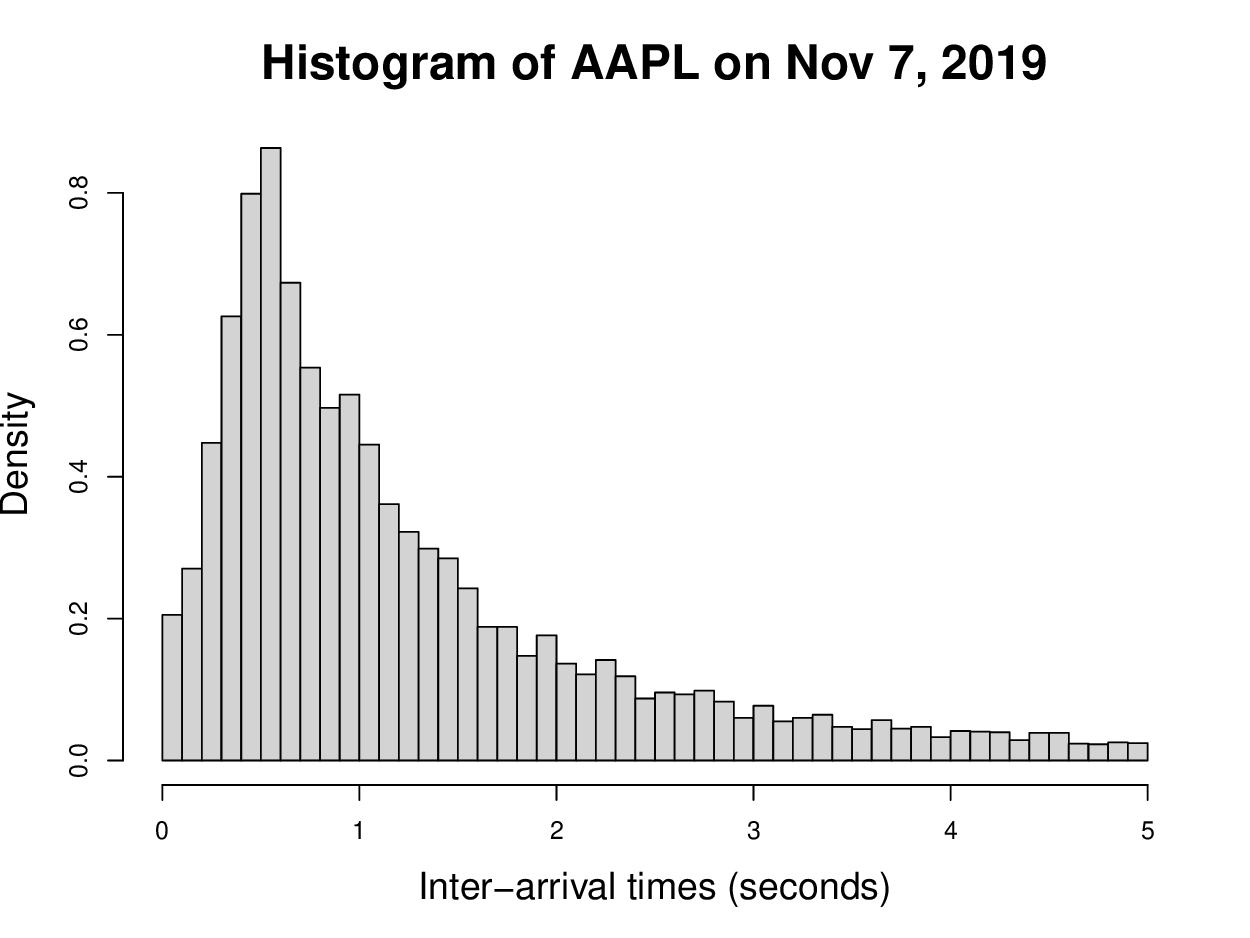}
		\caption{AAPL}\label{Fig:hist_AAPL_sp}
	\end{subfigure}
	\qquad
	\begin{subfigure}[b]{0.45\textwidth}
		\centering
		\includegraphics[width=\textwidth]{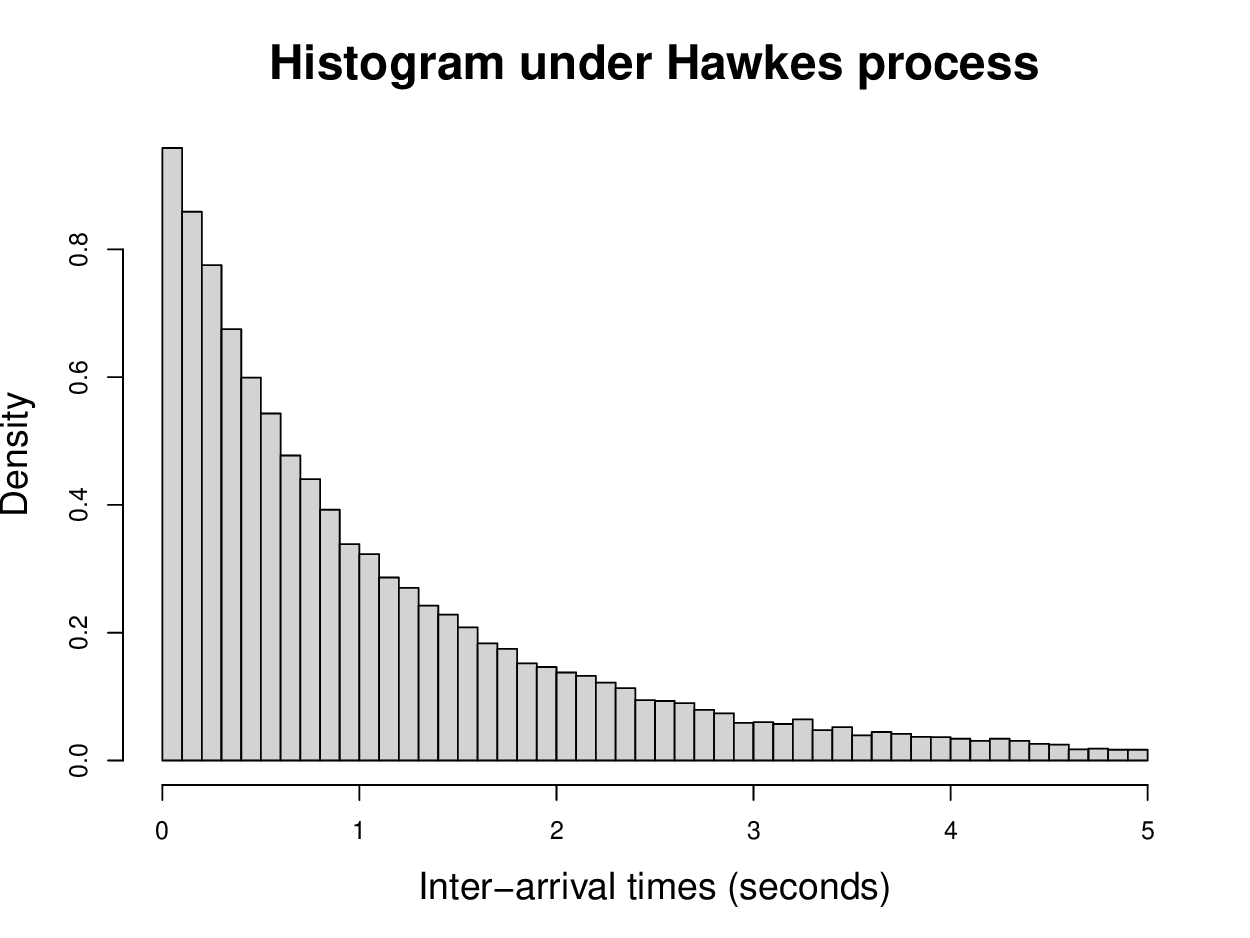}
		\caption{Hawkes process}\label{Fig:hist_Hawkes_sp}
	\end{subfigure}
	\\
	\begin{subfigure}[b]{0.45\textwidth}
		\centering
		\includegraphics[width=\textwidth]{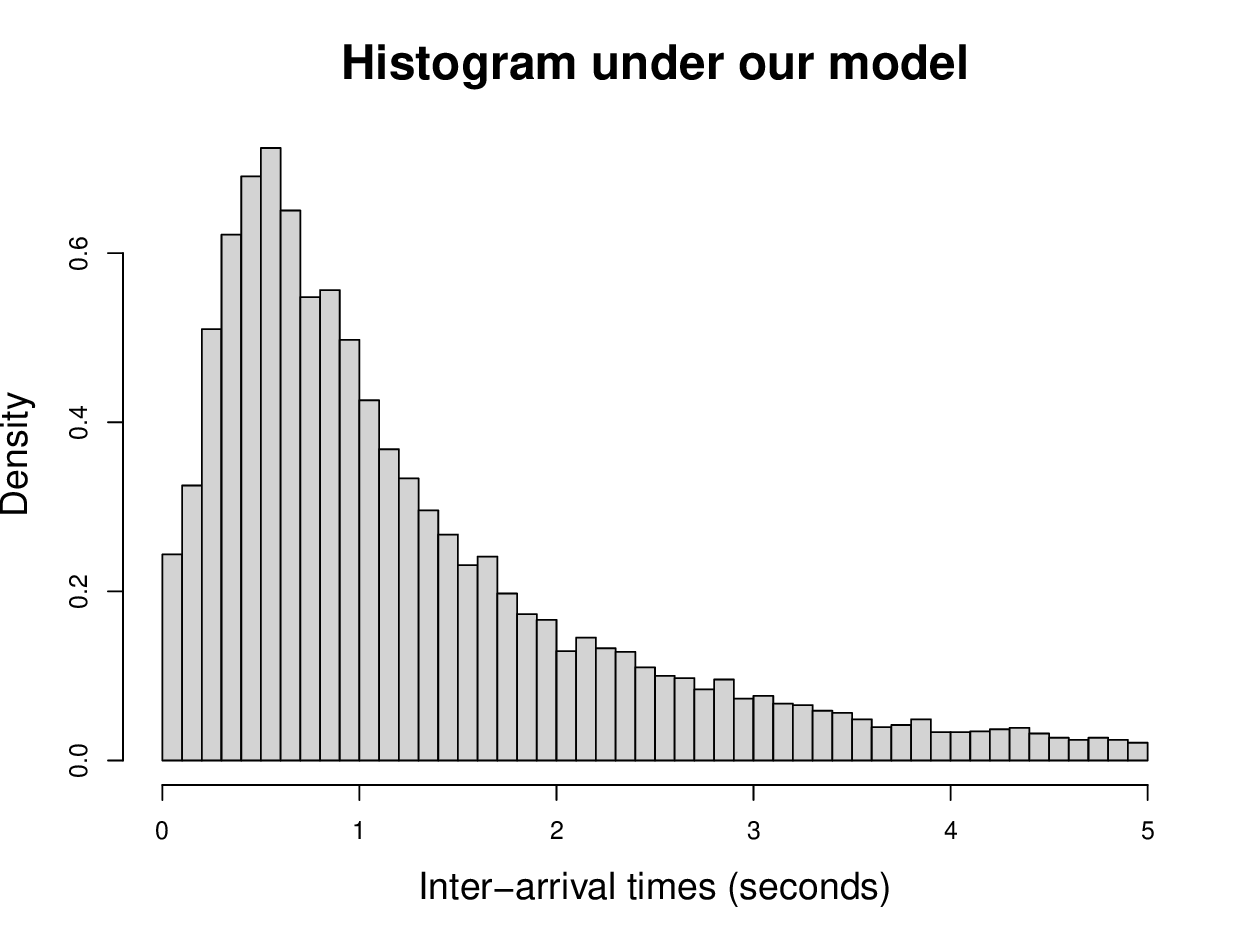}
		\caption{Our model}\label{Fig:hist_dHawkes_sp}
	\end{subfigure}
	\qquad
	\begin{subfigure}[b]{0.45\textwidth}
		\centering
		\includegraphics[width=\textwidth]{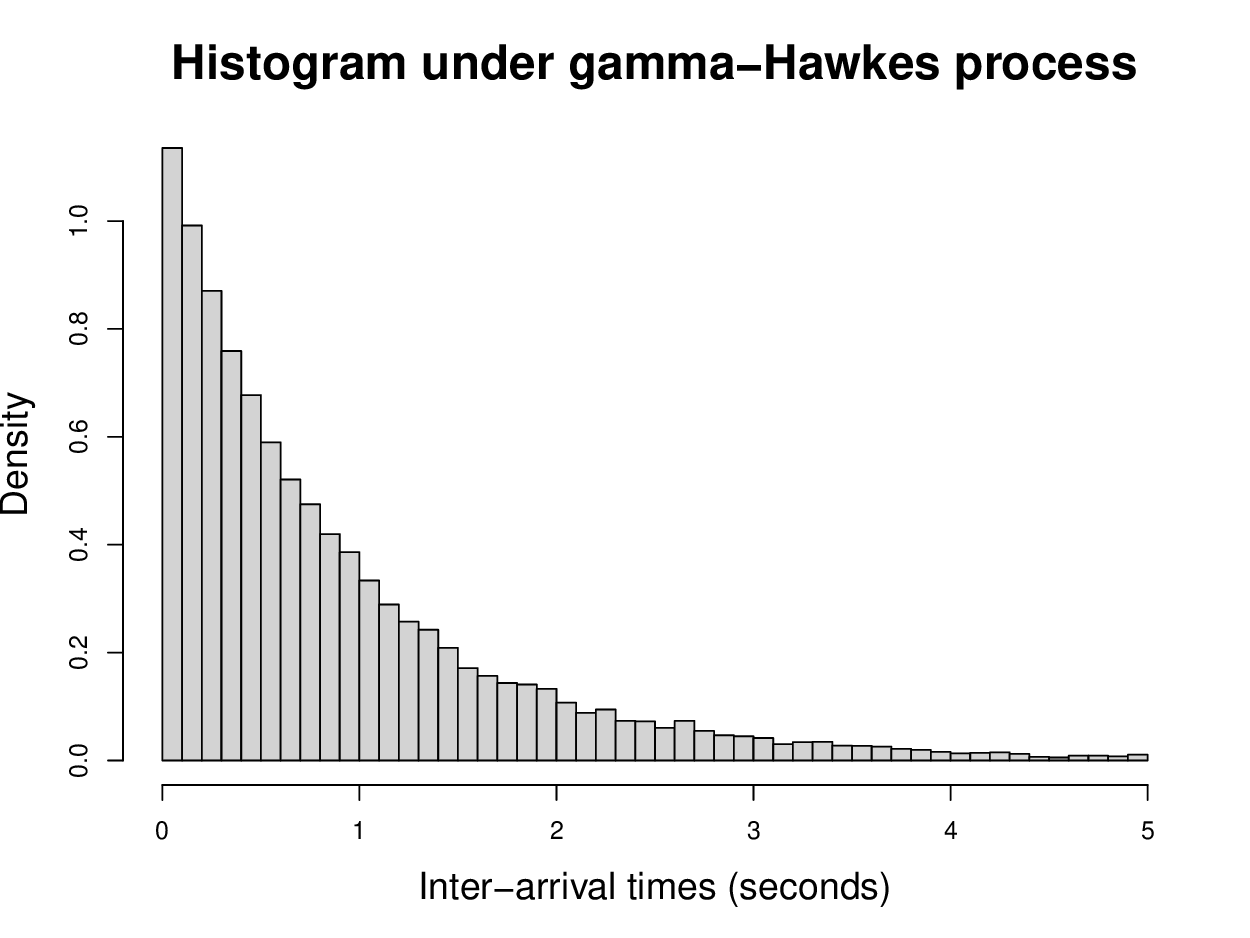}
		\caption{$\Gamma$-Hawkes process}\label{Fig:hist_gHawkes_sp}
	\end{subfigure}
	\caption{Histograms of inter-arrival times based on sparsely observed AAPL data with 0.5-second intervals, November 7, 2019 and simulated inter-arrivals under various Hawkes processes}
	\label{Fig:arrival_sparse}
\end{figure}

In the next step, we simulate arrival times using the inferred residuals and obtained estimates to examine the distribution.
The histogram in Figure~\ref{Fig:hist_AAPL_sp} illustrates the distribution of sparsely observed inter-arrival times for the AAPL mid-price on November 7, 2019.
Next, using the standard Hawkes model and the estimates from Eq.~\eqref{Eq:est_eHawkes}, the inter-arrival times are generated via the simulation and represented as a histogram in Figure~\ref{Fig:hist_Hawkes_sp}.
The generated inter-arrival times also display an exponential-like distribution due to the exponential distribution in the simulation. 

Meanwhile, the histogram in Figure~\ref{Fig:hist_dHawkes_sp}, depicting the inter-arrival times generated by the proposed model, displays a shape similar to the empirical distribution of inter-arrival times in Figure~\ref{Fig:hist_AAPL_sp}. 
The inter-arrival distribution fits well because the proposed model can flexibly adjust the residual distribution. 
This result is consistent with the findings presented in Subsection~\ref{Subsec:intraday}.

Next, the Hawkes process with the $\Gamma$-kernel \citep{Lesage} was applied for further comparison. 
The $\Gamma$-kernel is defined as follows:
\begin{equation}
	h(t) = \alpha \frac{(t \beta)^{k-1}\e^{-\beta t}}{\Gamma(k)} \label{Eq:gamma}
\end{equation}
where the parameter names are slightly modified to maintain consistency with the proposed model.
When $k=1$, Eq.~\eqref{Eq:gamma} reduces to the exponential kernel.
This model has a kernel structure combining power law and exponential components, 
which might suggest that it can generate versatile results.
Based on the sparsely observed data, the obtained MLE results are 
\begin{align*}
	&\hat \mu = 0.0915(0.0146), \quad \hat\alpha = 0.0436(0.0036), \quad \hat\beta =0.0475(0.0040)\\
	&\hat k =1.4367(0.0541).
\end{align*}
The $\Gamma$-Hawkes process is simulated using these estimates 
and Figure~\ref{Fig:hist_gHawkes_sp} presents the histogram of the generated inter-arrivals.
The histogram is quite similar to that generated under the standard Hawkes model,
and the proposed model better fits to the empirical inter-arrival distribution.

\subsubsection{Multivariate model and simulation}

The model is extended to multivariate frameworks and applied to sparsely observed data. Additionally, we employ various observation intervals to compare the standard Hawkes model with the proposed Hawkes model.
 
First, we fit the sparsely modified high-frequency price dynamics used in Subsection~\ref{subsubsec:oned} to the bivariate Hawkes model defined in Definition~\ref{Def:mHawkes}, 
with parameter constraints $\alpha_{11} = \alpha_{22}$ and $\alpha_{12} = \alpha_{21}$
for parsimony.
Although the bisymmetric kernel may appear simplified, 
it has been employed in empirical research under similar conditions, 
as demonstrated in studies by \cite{bacry2012non}, \cite{Bacry2014}, \cite{Barcry2015}, and \cite{lee2017modeling}.
To conserve space, we do not present it separately, but similar results were obtained without the constraint.

Second, we construct and estimate the proposed model, as defined in Definition~\ref{Def:mdH}, using the MLE described in Eq.~\eqref{Eq:MLE2} with the sparse data. 
The same parameter constraints as those applied to the standard Hawkes model are used.

Next, we perform a simulation and subsequently demonstrate its application by illustrating the daily volatility of stock price processes.
The price dynamics may follow a marked point process \citep{lee2017marked}, 
because the jump size (the mark) in the price dynamics may not be constant.
The mark is a random variable and may influecne future intensities.

For simplicity, the mark sizes are assumed to be independent without future influence on the $\lambda$ process.
Although it is feasible to implement a dependent marked model in which the mark size affects the $\lambda$ process in the proposed model, such an approach would increase the model complexity and is left for future research. 
Consequently, this paper focuses exclusively on independent mark distributions.

\begin{figure}[t]
	\begin{subfigure}{0.47\textwidth}
		\centering
		\includegraphics[width=\textwidth]{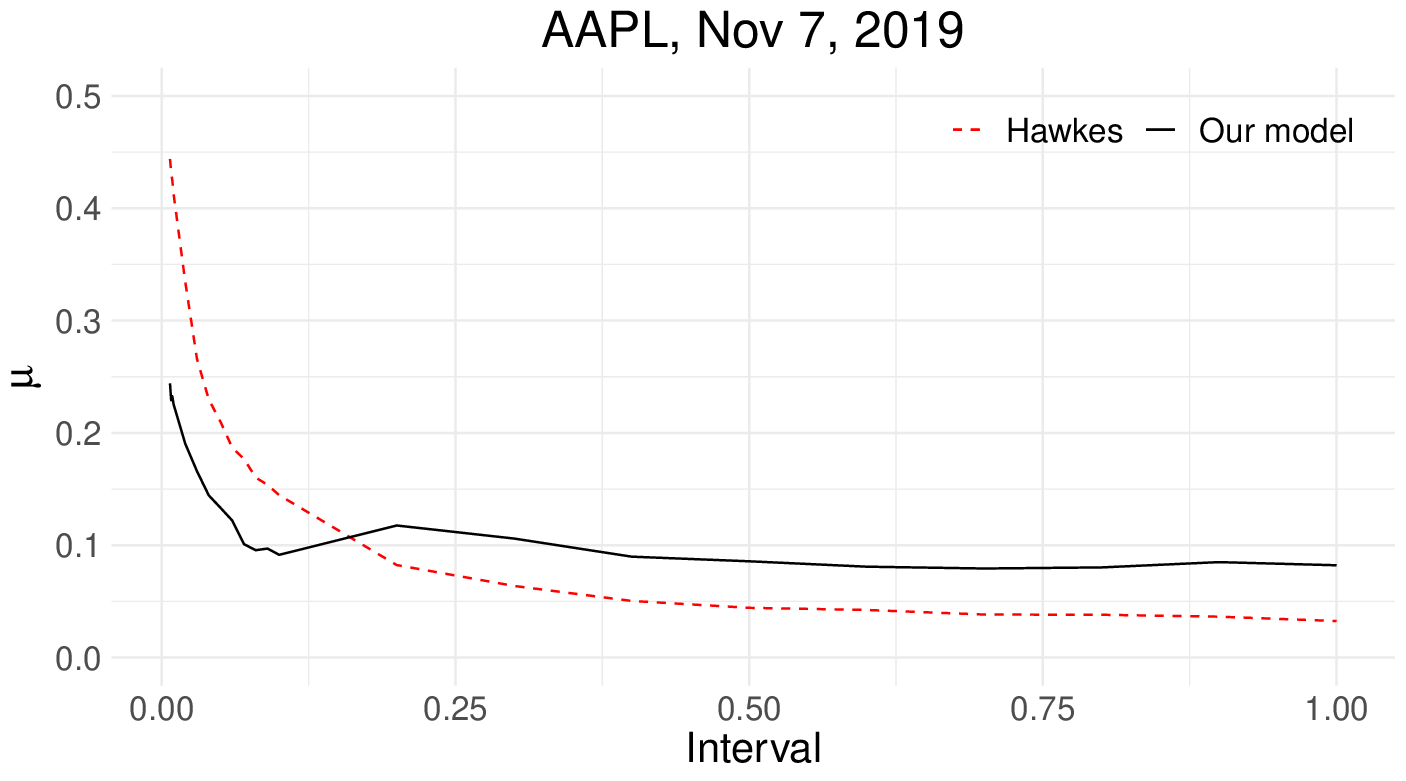}
		\caption{$\hat \mu$}
		\label{fig:mu}
	\end{subfigure}
	\quad
	\begin{subfigure}{0.47\textwidth}
		\centering
		\includegraphics[width=\textwidth]{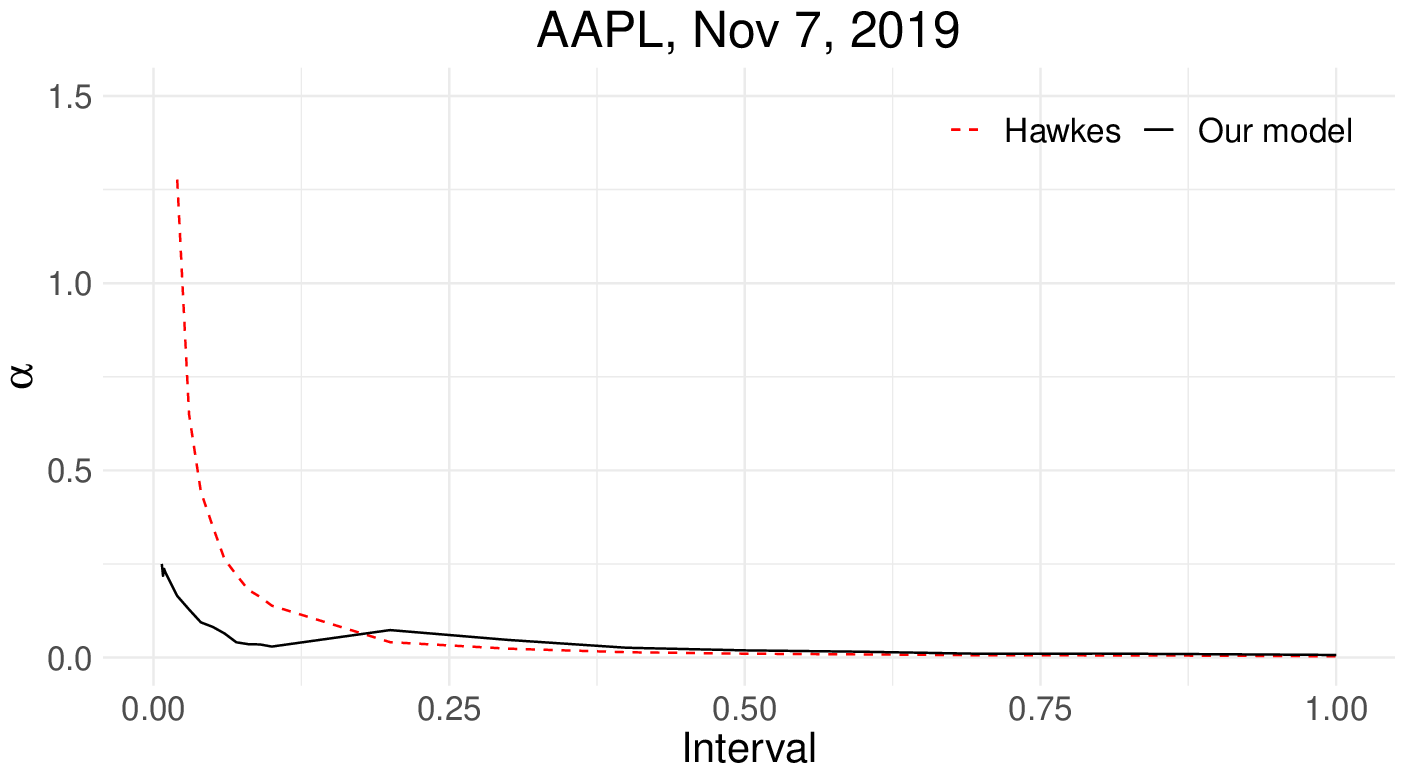}
		\caption{$\hat \alpha$}
		\label{fig:alpha}
	\end{subfigure}
	
	\vspace{5mm}
	
	\begin{subfigure}{0.47\textwidth}
		\centering
		\includegraphics[width=\textwidth]{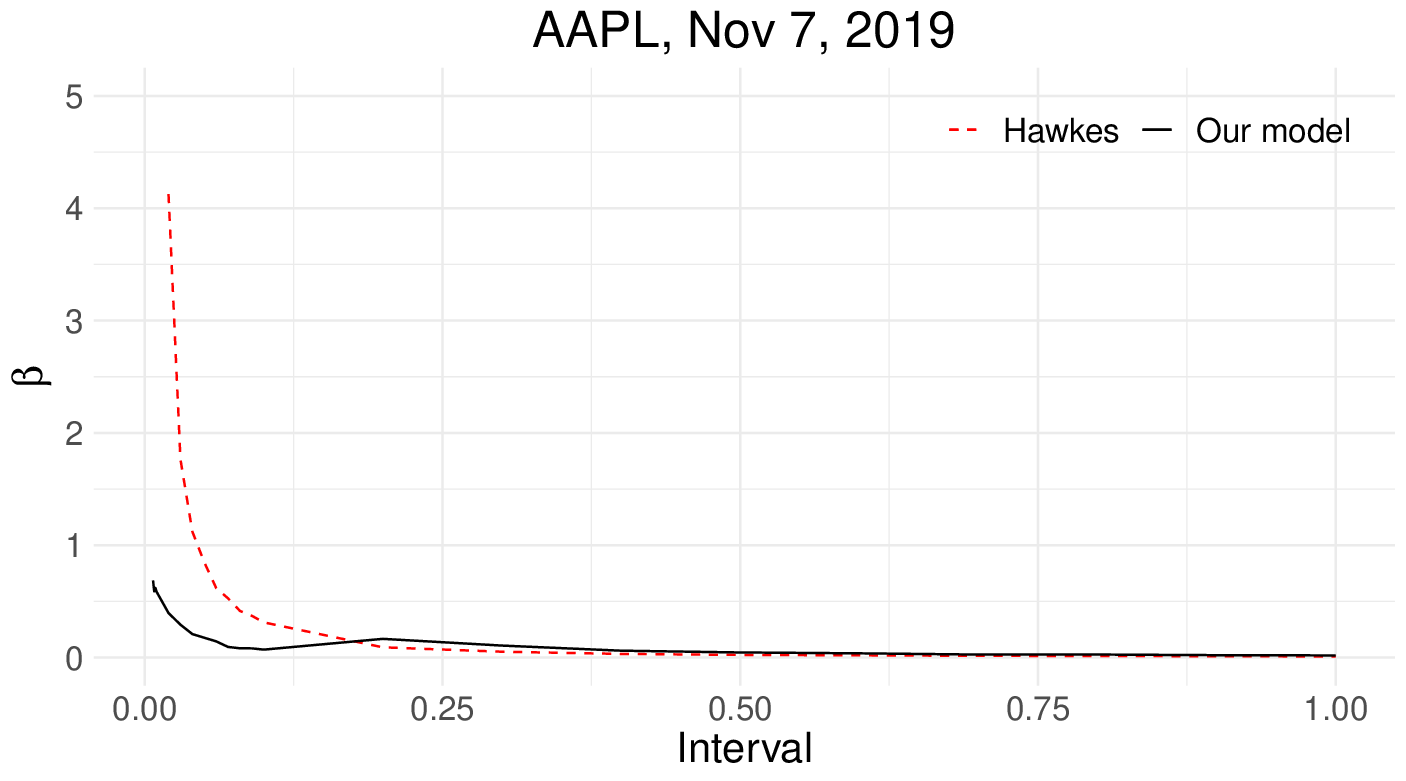}
		\caption{$\hat \beta$}
		\label{fig:beta}
	\end{subfigure}
	\quad
	\begin{subfigure}{0.47\textwidth}
		\centering
		\includegraphics[width=\textwidth]{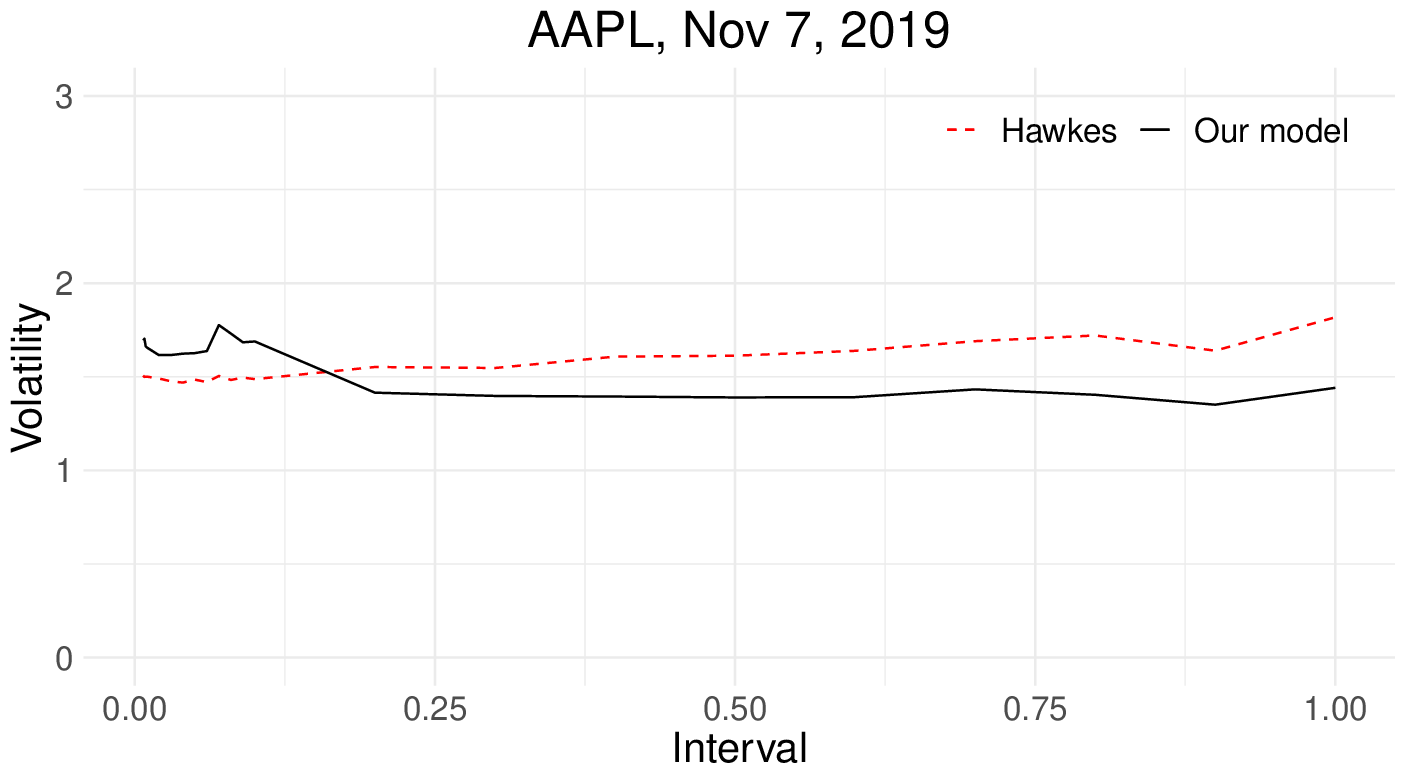}
		\caption{Daily volatility}
		\label{fig:vol}
	\end{subfigure}
	\caption{Various estimates and estimated daily price volatility across various observation intervals in data preprocessing}~\label{Fig:compare}
\end{figure}

With the estimated parameters, 
$M=10,000$ paths of price dynamics are generated using Algorithm~\ref{Algo:2}
applying the same time horizon as the original data.
During the simulation, the jump size in price change is also bootstrapped 
from the set of realized jumps from the sparsely modified data.

Therefore, we generate $M$ simulated the bivariate processes for the up and down movements, along with the marks.
The sample standard deviation of the difference between the up and down price processes at the end of the time horizon $[0, t]$ is computed as follows:
$$ \textrm{S.D.}(N_1(t) - N_2(t)) $$
where $N_1$ and $N_2$ denotes the jump-size-weighted number of the up and down movements, respectively.

We compare the results with the volatility calculated using the standard Hawkes model.
After estimating the Hawkes model, we compute the standard deviation of the daily price change with Eq.~\eqref{Eq:Hvol}.

The estimates of the Hawkes model from the estimation results in Figure~\ref{Fig:compare} vary more than those of the proposed model according to the interval, particularly for $\hat \alpha$ and $\hat \beta$, which is especially noticeable when the interval is short. 
As the observation interval approaches zero, it converges to the case with using raw data discussed in Subsection~\ref{Subsec:intraday}.

Meanwhile, in our model, $\hat \alpha$ and $\hat \beta$ are relatively constant over the interval,
indicating more robust estimations.
The estimates associated with $\mu$ display a similar trend,
because the estimates of the standard Hawkes process tend to be large as the observation interval approaches zero, implying ultra-high-frequency activities, a high excitation, and fast decay. 
However, in the proposed process, the fat-tailed residuals observed in Subsection~\ref{Subsec:intraday} can manage these extreme activities, resulting in relatively unchanged estimates.

Both the Hawkes volatility, based on Eq.~\eqref{Eq:Hvol}, and volatility based on our model, demonstrate stable estimates over intervals.
This result suggests that the sparse observation method, along with the Hawkes volatility derived from it, is expected to have a significant degree of reliability.

\section{Conclusion}~\label{Sect:concl}

This study introduces a point process model with a flexible residual distribution, characterized by excitation features and a discrete Markovian structure. 
We demonstrate how flexible residual modeling is directly related to the intensity model.
This flexibility facilitates efficient estimation and simulation while preserving the original properties of the stochastic processes.
The usefulness of the proposed model was examined via examples using high-frequency financial data.
This study demonstrates that the flexible model fits distributions more accurately and provides stable estimation results regardless of the data preprocessing, compared to the typical Hawkes model.
This finding suggests that a broader range of research opportunities can be anticipated in the future using our model.

%\bmhead{Supplementary information}

\begin{appendices}

\section{Hawkes volatility formula}\label{App:vol}
We assume that jump size in price dynamics is independent of the previous information.
Thus, the Hawkes volatility under the model in Definition~\ref{Def:mHawkes} is calculated as follows:
\begin{equation}
	\mathrm{Hvol}_t= \sqrt{ \bm{\mathrm{u}}^{\top} \left[ \overline \bZ \circ \bB  +  (\overline \bZ \circ \bB )^{\top} +  \overline \bZ^{(2)}\circ \Dg (\E [\bm{\lambda}_t]) \right] \bm{\mathrm{u}} t } \label{Eq:Hvol}
\end{equation}
where
$\bm{\mathrm{u}} = \begin{bmatrix} 1 & -1 \end{bmatrix}^{\top}$
and
\[ 
\overline \bZ = \E \begin{bmatrix} z_1 & z_2 \\  z_1 & z_2 \end{bmatrix},
\quad
\overline \bZ^{(2)} = \E \begin{bmatrix} z_1^2 & z_2^2 \\  z_1^2 & z_2^2 \end{bmatrix}
\]
with $z_i$ representing the jump size in price dynamics of type $i$ and
$\mathrm{Dg}$ denoting a diagonal matrix.
In addition, $\bB$ denotes a $2\times 2$ matrix that satisfies the following:
\begin{equation}
	\bB(\bm{\alpha} - \bm{\beta})^{\top} + \overline \bZ^{\top} \circ \E[\bm{\lambda}_t\bm{\lambda}_t^{\top}] +\Dg(\E[\bm{\lambda}_t]) \left(\bm{\alpha}  \circ \overline \bZ  \right)^{\top} - \Dg(\overline \bZ) \E[\bm{\lambda}_t] \E[\bm{\lambda}_t]^{\top} = \bm{0},
\end{equation}
where
$\E[\bm{\lambda}_t\bm{\lambda}_t^{\top}]$ indicates a $2\times 2$ matrix that satisfies
\begin{equation}	
	(\bm{\alpha} - \bm{\beta}) \E[\bm{\lambda}_t \bm{\lambda}_t^{\top} ] + \E[\bm{\lambda}_t \bm{\lambda}_t^{\top} ] (\bm{\alpha} - \bm{\beta})^{\top} + \bm{\alpha} \Dg(\E[\bm{\lambda}_t]) \bm{\alpha}^{\top} = \bm{0},
\end{equation}
and $\E[\bm{\lambda}_t]$ represents a $2\times 1$ matrix that satisfies
\begin{equation}
	\E[\bm{\lambda}_t] = ( \bm{\beta} - \bm{\alpha})^{-1}\bm{\beta}\bm{\mu}\label{Eq:E_lambda2}
\end{equation}
with
\[
\bm{\mu} = \begin{bmatrix} \mu_1 \\ \mu_2 \end{bmatrix}, \quad  
\bm{\alpha} = \begin{bmatrix} \alpha_{11} & \alpha_{12} \\ \alpha_{12} & \alpha_{11} \end{bmatrix}, \quad \bm{\beta} = \begin{bmatrix} \beta_{1} & 0 \\ 0 & \beta_{2} \end{bmatrix}.
\]
For more detailed information, please refer to the work by~\cite{LeeHawkesVol}.

\end{appendices}

%%===========================================================================================%%
%% If you are submitting to one of the Nature Portfolio journals, using the eJP submission   %%
%% system, please include the references within the manuscript file itself. You may do this  %%
%% by copying the reference list from your .bbl file, paste it into the main manuscript .tex %%
%% file, and delete the associated \verb+\bibliography+ commands.                            %%
%%===========================================================================================%%

\bibliography{Bib}
\bibliographystyle{chicago}

\end{document}